\documentclass[12pt,preprint]{aastex}

\def\1h{{1H 0419-577}}

\def\msun{{\rm M_{\odot}}}

\def\xmm{{\it XMM-Newton}}
\def\chandra{{\it Chandra}}

\def\et{{et al.\ }}

\def\asca{{\it ASCA}}
\def\sax{{\it BeppoSAX}}

\def\arcs{{\hbox{$^{\prime\prime}$}}}

\def\H0{{\rm ~km~s^{-1}~Mpc^{-1}}}

\def\msun{M_{\rm \odot}}

\def\et{{et al.}}

\shorttitle{XMM-Newton observation of 1H0419-577}
\shortauthors{Pounds et al.}

\begin{document}


\title{Resolving the large scale spectral variability of the luminous Seyfert 1 galaxy \1h: Evidence for a new 
emission component
and absorption by cold dense matter.}


\author{K.A.Pounds\altaffilmark{1}, J.N.Reeves\altaffilmark{2,3}, K.L.Page\altaffilmark{1},  
P.T.O'Brien\altaffilmark{1}}

\altaffiltext{1}{Department of Physics and Astronomy, University of Leicester, Leicester LE1 7RH, UK}
\altaffiltext{2}{Laboratory for High Energy Astrophysics, Code 662, NASA Goddard Space Flight Center,
Greenbelt, MD 20771, USA}
\altaffiltext{3}{Universities Space Research Association}
\email{kap@star.le.ac.uk}


\begin{abstract}

An \xmm\ observation of the luminous Seyfert 1 galaxy \1h\ in September 2002, when the source was in an extreme
low-flux state, found a very hard X-ray spectrum at 1--10 keV with a strong soft excess below $\sim$1 keV.
Comparison with an earlier \xmm\ observation when \1h\ was `X-ray bright' indicated the dominant spectral
variability was due to a steep power law or cool Comptonised thermal emission. Four further \xmm\ observations,
with \1h\ in intermediate flux states, now support that conclusion, while we also find the variable emission
component in intermediate state difference spectra to be strongly modified by absorption in low ionisation matter.
The variable `soft excess' then appears to be an artefact of absorption of the underlying continuum while the
`core' soft emission can be attributed to recombination in an extended region of more highly ionised gas. We note
the wider implications of finding substantial cold dense matter overlying (or embedded in) the X-ray continuum
source in a luminous Seyfert 1 galaxy.

\end{abstract}

\keywords{X-ray astronomy:XMM-Newton:Seyfert galaxies:1H0419-577,LB 1727}

\section{Introduction}

\1h (also known as LB 1727) is a radio-quiet (8.4 GHz flux $\sim$3 mJy; Brissenden \et 1987) Seyfert galaxy at a
redshift z$\sim$0.104.  Optical spectra from the AAT (Turner \et\
1999) and ESO (Guainazzi \et\ 1998) showed \1h\ to be a typical broad line Seyfert 1 with a strong Big Blue Bump
(BBB). It has been widely studied at X-ray energies and found to exhibit an unusual degree of spectral
variability. A short pn-camera observation early in the \xmm\ programme reported a `typical' Seyfert 1  X-ray
spectrum with a power law of photon index $\Gamma$$\sim$1.9 together with a strong soft excess (Page \et\ 2002).
Using the  H$\beta$ line width from Grupe \et\ (2004) and the relationship with black hole mass of Kaspi \et\
(2000), we estimate a supermassive black hole (SMBH) of $1.3\times10^{8}$$\msun$. The [OIII] 5007\AA\ line
width measure of the stellar  velocity dispersion (Nelson 2000) gives a very similar figure.

To improve the X-ray data on \1h\ a new series of 5 \xmm\ observations was carried out, at approximately 3-monthly
intervals over the period September 2002 to September 2003. The first of those new observations, when \1h\ was
found to be in an extremely faint state, has been reported in Pounds \et\ (2004); hereafter Paper 1.  Three points
of particular note resulting from that first new \xmm\ observation of \1h\ were:  (1) an unusually hard (flat)
power law fit ($\Gamma$$\sim$1.0) to the EPIC data over the 2--10 keV band also exhibited curvature indicative of
an extreme relativistic Fe K emission line or partial covering of the continuum by a substantial column of `cold'
gas;  (2) although the RGS features were faint, the unambiguous detection of emission lines of OVII and OVIII
provided clear evidence for an extended region of photo-ionised gas in the nucleus of \1h; (3) a comparison of the
raw EPIC data with data obtained in December 2000, when \1h\ was considerably brighter in the X-ray band, showed
that the large-scale spectral variability in could be described by a variable, steep ($\Gamma$$\sim$2.5)
power law component.

In the present paper we analyse the remaining 4 observations from the new \xmm\ campaign (in December 2002, March,
June and September 2003), which together explore the X-ray spectrum of \1h\ over a range of flux states lying
between the extremes of December 2000 and September 2002.

\section{Observations and X-ray data}

The new observations took place on 27 December 2002 (orbit 558), 30 March (orbit 605), 25 June (orbit 649) and 16
September (orbit 690) in 2003, with on-target exposures of $\sim$10.2 ksec, $\sim$13.9 ksec, $\sim$13.1 ksec and
$\sim$13.9 ksec, respectively. X-ray data were available throughout each observation from the EPIC pn (Str\"{u}der
\et\ 2001) and MOS (Turner \et\ 2001) cameras, and the Reflection Grating Spectrometer/RGS (den Herder \et\ 2001).
These new data add to the earlier \xmm\ observations of \1h\ on 4 December 2000 (orbit 181) and 5 September 2002
(orbit 512) when the source was found to be in historically high and low flux states, respectively. Those earlier
observations have been reported in Page \et\ (2002) and in  Paper 1.

EPIC data were initially screened with the XMM SAS v5.4 software and events corresponding to patterns 0-4 (single
and double pixel events) were selected for the pn data and patterns 0-12 for MOS1 and MOS2, the latter then being
combined. A low energy cut of 300 eV was applied to all EPIC data and known hot or bad pixels were removed. Source
counts were obtained from a circular region of 45\arcs\ radius centred on \1h, with the background being taken
from a similar region offset from, but close to, the source. The X-ray light curve of \1h\ was essentially flat
throughout each observation and the background rate was low (except in orbit 558, when noisy data were edited
out). We therefore integrated each data set for spectral analysis. Individual EPIC spectra were binned to a
minimum of 100 counts per energy bin to facilitate use of the $\chi^2$ minimalisation technique in spectral
fitting and ensure adequate statistics in the  5--10 keV band. Spectral fitting was based on the Xspec package
(Arnaud 1996) and all fits included absorption due to the line-of-sight Galactic column
$N_{H}=2\times10^{20}\rm{cm}^{-2}$. Errors are quoted at the 90\% confidence level ($\Delta \chi^{2}=2.7$ for one
interesting parameter).

We were fortunate to observe \1h\ in widely differing flux states, ranging conveniently between the low and high
flux state extremes reported previously. Figure 1 shows the background-subtracted (pn camera) spectral data for all
6 observations, compared with a mean power law fit (at 2--10 keV) of $\Gamma$=1.73.  The plot shows very clearly
that the largest flux variation occurs in the soft X-ray band, with all spectra converging above $\sim$5 keV. The
only significance we would claim for the mean power law  fit is in the spectral index being close to the
`canonical' value for Seyfert 1 galaxies (for fits excluding reflection, Nandra and Pounds 1994), suggesting that 
this unusually variable
AGN has an $\it average$ X-ray spectrum typical of the class. Retaining our previous descriptions of the orbit 181
and 512 data as, respectively, `high' and `low' flux states, we now define orbit 558 as `mid-low', orbit 605 as
`mid-high' and group orbits 649 and 690 together to give a well-defined `intermediate' flux state spectrum. 

In the following analysis spectral parameters are quoted in the rest frame of \1h, while figures (except fig.10)
retain the observed photon energy scale. In general we have fitted the pn and MOS data simultaneously, with only
the power law parameters untied, reflecting the well-known systematic difference in deduced continuum slopes.

\section{Spectral fitting of the intermediate flux state EPIC data} 

We began the EPIC spectral analysis in the conventional way by fitting a power law over the hard X-ray (2--10 keV)
band, aiming thereby to minimise the effects of soft X-ray emission and/or low energy absorption. This fit yielded
a photon index of $\Gamma$=1.60$\pm$0.01 (pn) and $\Gamma$=1.55$\pm$0.02 (MOS). Statistically, the simple power
law fit was quite good, with $\chi^{2}$ of 393 for 357 degrees of  freedom (dof). The most obvious residual
spectral feature is a weak emission line observed just below $\sim$6 keV (figure 2a). The addition of a gaussian
emission line improved the fit (to $\chi^{2}$ = 373/353 dof), with a line energy (in the AGN rest-frame) of
6.25$\pm$0.12 keV, rms width $\sigma$=320$\pm$150 eV, flux = 1.5$\pm$0.6$\times10^{-5}$~ph s$^{-1}$ cm$^{-2}$, and
EW = 100$\pm$40 eV. 

Extending this spectral fit to 0.3 keV revealed the strong excess below $\sim$1 keV (figure 2b). The addition of
blackbody components of kT $\sim$110 eV and kT $\sim$250 eV modelled the soft excess quite well, but it was
necessary to add (ad hoc) absorption edges at $\sim$0.62 keV ($\tau$$\sim$0.22), $\sim$0.74 keV ($\tau$$\sim$0.26)
and $\sim$0.90 keV ($\tau$$\sim$0.10) to achieve a statistically acceptable fit  ($\chi^{2}$ = 825/783 dof).
Figure 3 reproduces this `conventional' fit to the intermediate flux  level data for \1h, which we note has model
parameters typical for a luminous Seyfert 1 galaxy.

The above spectral fitting allowed the mean X-ray fluxes of the intermediate state spectrum of \1h\ to be determined.
These were: $5.9\times10^{-12}$~erg s$^{-1}$ cm$^{-2}$ (0.3--1 keV), $3.0\times10^{-12}$~erg s$^{-1}$ cm$^{-2}$ 
(1--2 keV), and $1.05\times10^{-11}$~erg s$^{-1}$ cm$^{-2}$ (2--10 keV). Combining these fluxes yields an
`intermediate state' 0.3--10 keV luminosity for \1h\ of $4.3\times 10^{44}$~erg s$^{-1}$ ($ H_0 = 75
$~km\,s$^{-1}$\,Mpc$^{-1}$).

\section{An alternative spectral fit to the intermediate flux state spectrum}

Our main aim in the present study is to better understand the large scale spectral variability of \1h\ - and
thereby of AGN more generally.  In Paper 1 we found that the difference spectrum between high and low
state EPIC pn data (the high flux difference spectrum) could be well fitted by a power law of $\Gamma$$\sim$2.5,
steepening to $\sim$2.7 above $\sim$2 keV. This simple fit suggested that the  low state spectrum (hard power law
plus soft excess) might represent a non-varying `core' emission, leaving the main spectral change to be
represented by the variable flux power law component. To further explore that possibility we now re-analyse the
new intermediate flux state EPIC data, after subtracting the September 2002 data, to yield the intermediate state
difference spectrum. 

We find the resulting intermediate state difference spectrum is similar to the high state difference spectrum,
with a mean 0.3-10 keV power law index of $\Gamma$$\sim$2.41 (pn), steepening to $\sim$2.85 when fitted above
$\sim$2 keV.  However, the fit is much less good ($\chi^{2}$ = 1174/799 dof), due to a broad deficit of flux at
$\sim$0.5--1 keV (figure 4a). If the large-scale spectral variability in \1h\ is indeed well modelled by a steep,
variable flux power law, as proposed in our previous analysis of the high state difference spectrum (Paper 1), the
new data suggest the variable emission component is modified by absorption in intermediate flux states. To model
that possibility we then compared the intermediate state difference spectrum with a power law plus a photoionised
absorber, represented by XSTAR (Kallman \et\ 1996). Here, the absorption is compared with a grid of ionised
absorbers, with column density, ionisation parameter $\xi$(= $L/nr^2$, where n is the gas density at a distance r
from the ionising source of luminosity L) and outflow (or inflow) velocity as variable parameters. All abundant
elements from C to Fe are included, with the relative abundances as further  variable parameters. The resulting
fit was good ($\chi^{2}$ of 839/790), with a column density $N_{H}$$\sim$$4\times10^{21}\rm{cm}^{-2}$ of low
ionisation gas (log$\xi$=-1.7$\pm$0.4).  The  relative abundances of the key elements, of C, N, O, Ne, Mg and Fe,
were 0.7, 0.9, 0.20, 0.75, 1.0 and 1.0, though only O, Ne and Fe were well constrained in the fit. This best fit was
obtained with a redshift of 0.1$\pm$0.005, implying the substantial column of `cold' absorbing gas is local to \1h.
The power law slopes in the fit increased to $\Gamma$$\sim$2.8 (pn) and $\Gamma$$\sim$2.7 (MOS), while still
leaving a further spectral steepening above $\sim$3 keV (Figure 4c). Figure 4b illustrates the relevant XSTAR
model, where the absorption structure is dominated by continuum absorption, in increasing photon energy, of C, O, 
Fe and Ne.

\subsection{The form of the variable emission component}

In the above spectral fit we assumed the variable emission component in \1h\ has the form of a power law. A
similar conclusion was  proposed from  an \asca\ study of the Seyfert 1 galaxy MCG-6-30-15 (Shih \et\ 2002), and
supported by extended \xmm\ observations of the same source (Fabian and Vaughan 2003).  However, it is notable
that for \1h\ the power law fit including absorption (figure 4c) indicates further steepening above $\sim$3 keV,
and we recall the compTT model (Titarchuk 1994) gave an even better fit to the high state difference spectrum
reported in Paper 1. Figure 5 reproduces the residuals to the single power law and a compTT fits to the high state
difference spectrum. The thermal Comptonisation  model has the additional appeal of being more physical than a
power law fit. Since the intrinsic curvature of the thermal continuum might significantly affect the derived
absorption parameters we therefore repeated the analysis of the intermediate state difference spectrum with a
model involving the emergence of a cool Comptonised emission component modified by absorption in ionised
matter. 

Replacing the power law of section 4 with a Comptonised emission component, with an initial temperature of
kT$\sim$2.7 keV and optical depth  $\sim$4.4, as found for the high flux difference spectrum, the fit was indeed
better than for the power law model,  though still poor ($\chi^{2}$ = 1102/796 dof), with data:model residuals
similar to those for the power law fit in figure 4a.  Photoionised absorption represented by XSTAR was then added
to the model. The outcome was a very good fit ($\chi^{2}$ = 828/784 dof) for an Comptonised emission  component of
kT=2.3$\pm$0.4 keV and optical depth 4.5$\pm$0.7. The absorption was again well modelled by low ionisation matter,
with a column density of $N_{H}$$\sim$$4.4\times10^{21}\rm{cm}^{-2}$ and ionisation parameter of
log$\xi$=-1.8$\pm$0.3. The relative abundances of the key elements, of C,N,O,Ne,Mg and Fe, were
0.2,0.2,0.12,0.4,0.7 and 0.66, with the lower abundances of C and N (compared with the power law fit) adjusting to
the low energy curvature in the continuum fit.

Figure 6a shows the best-fit Comptonised emission component and pn camera data for the intermediate flux
difference spectrum, with the XSTAR absorption component removed. Figure 6b reproduces the model spectrum, while
figure 6c illustrates the quality of the resulting fit.

\section{Spectral fits to the mid-low flux state spectrum}

Taken together with the high state difference spectrum reported in Paper 1, the above analysis of the intermediate
state difference spectrum suggests the large scale spectral variability of \1h\ is indeed due to an emerging
emission component which - at intermediate flux levels - bears the imprint of absorption by low ionisation matter.
To see if that trend of variable absorption with flux level is continous, we repeated the above analysis for the
mid-low flux state observation, but again starting with a `conventional' power law plus blackbody fit.

Fitting a power law over the hard X-ray (2--10 keV) band yielded a photon index of $\Gamma$=1.41$\pm$0.03 (pn) and
$\Gamma$=1.35$\pm$0.06 (MOS) for the mid-low flux spectrum. Extending this fit to 0.3 keV showed a soft excess rising 
sharply below $\sim$0.7 keV (figure 7a). The narrower profile of this soft excess (compared with that in figure 2a)
allowed it to be well-modelled by the addition of a single blackbody component of kT $\sim$102 eV, no hotter blackbody
being required. In this case a single absorption edge ($\tau$$\sim$0.8 at 0.76$\pm$0.01 keV) was needed to complete an 
excellent fit ($\chi^{2}$ = 245/244 dof). Figure 7b shows the ratio of data to this power law, blackbody and absorption
edge model. In summary, a conventional fit to the mid-low flux spectrum of \1h\ shows a hard (flat) power law, with a
sharply rising (cool) soft excess, and a deep absorption edge at $\sim$0.76 keV (in the AGN rest frame).

The above fit provided a measure of the X-ray fluxes and luminosity of \1h\ in the the mid-low flux state, which
were: $3.3\times10^{-12}$~erg s$^{-1}$ cm$^{-2}$ (0.3--1 keV),  $1.5\times10^{-12}$~erg s$^{-1}$ cm$^{-2}$  (1--2
keV), and $8.3\times10^{-12}$~erg s$^{-1}$ cm$^{-2}$ (2--10 keV), corresponding to a 0.3--10 keV luminosity for
\1h\ of $2.8\times 10^{44}$~erg s$^{-1}$ ($ H_0 = 75 $~km\,s$^{-1}$\,Mpc$^{-1}$).

Proceeding to examine the mid-low state difference spectrum, we followed the analysis procedure described in
section 4, using compTT to model the emission component. The relatively poor statistics meant that the
Comptonisation parameters were not well defined, the best fit having a temperature kT=2.1$\pm$1.5 keV and optical
depth 2.7$\pm$2. However, the
continuum fit was adequate to show the absorption trough to be noticeably deeper in the mid-low state difference
spectrum (figure 8a). Fitting this absorption with XSTAR, with abundances fixed at the values found in the
intermediate state model, again produced an acceptable fit ($\chi^{2}$ = 259/238 dof), with a column density of
$N_{H}$$\sim$$2\times10^{22}\rm{cm}^{-2}$ and ionisation parameter of log$\xi$=-2.3$\pm$0.3. While the ionisation
parameters of the intermediate and mid-low difference spectra are the same within the formal errors, we see below
that the lower ionisation in the mid-low flux state spectrum is critical to understanding the different shape of
the absorption trough in the two spectra. Figure 8b reproduces the XSTAR model and figure 8c shows the data:model
residuals for the mid-low difference spectrum fit.

\section{The form of the variable absorption}

The above difference spectrum fits indicate that absorption in low ionisation matter is substantially modifying
the variable emission component - and hence the overall observed  X-ray spectrum of \1h; furthermore, we find the
absorbing matter to become more ionised and the column density to fall as the continuum flux rises. To further
explore the variable absorber we then computed the ratio of the `raw' intermediate and mid-low flux spectra.  The
shape of that ratio plot (figure 9a) is particularly interesting, the marked drop observed at $\sim$0.7 keV
coinciding with low ionisation Fe absorption edges, with no obvious change in the flux ratio at the corresponding
edges of low ionisation oxygen. The flux ratio plot is thus consistent with the specific XSTAR models which match
the relatively deeper absorption edge near 0.7 keV with the lower ionisation parameter for the lower flux state
spectrum (cf figures 6b and 8b). A weaker feature at $\sim$0.9 keV in figure 9a, close to the absorption edge
structure of Ne, is also qualitatively consistent with the individual XSTAR fits.  Figure 9b shows the ratio of
the mid-high to high flux state spectra, indicating the same pattern of decreasing absorption continues as
the flux level of \1h\ rises towards the high state.

The different profile of the absorption in the intermediate and mid-low state difference spectral fits is determined 
by the changing ionisation parameter. To better understand that change we replot
the key section of each XSTAR model fit in figure 10, where for ease of comparison with listed edges the energy
axes are adjusted to the rest-frame of \1h and the principal absorbing ion stages are noted against the respective
absorption features. Reference to the detailed absorption cross-sections for ground states of O and Fe (Kallman
and Bautista 2001) then provides a qualitative explanation for the differential absorption we see in \1h, since
whereas the  threshold energy cross-section in Fe increases by a factor $\sim$2.5 from FeV-FeI, the threshold
cross-sections increases by only $\sim$40  percent from OIV-OI. 

In summary, we find that the emerging emission component responsible for the main spectral change in \1h\ is
modified by low energy absorption in a substantial column density of low ionisation matter. Further, while
remaining low, the ionisation state of the absorber increases as \1h\ gets brighter, while the absorbing
column (or perhaps the covering factor) simultaneously decreases, and the increasing ionisation parameter provides
a natural explanation of the changing energy profile of the absorption trough.

\section{Spectral lines in the RGS data}

Given the above evidence for substantial absorbing matter modifying the EPIC spectrum, it is of obvious importance
to check whether this is consistent with the higher resolution RGS data. To pursue that question we examined the
simultaneous \xmm\ grating data of \1h, initially summing the data from all 5 orbits (512-690) to get the best
statistics, for what is a relatively faint source. Figure 11 displays the RGS-1 and RGS-2 fluxed spectrum,
binned relatively coarsely at 85 m\AA. The only spectral lines detected are all in emission, though the broad
deficit between $\sim$10--20 \AA\ is qualitatively consistent with the absorption trough seen in the EPIC spectra.
The narrower absorption feature observed near 17.5 \AA\ is also consistent with the Fe 2-3 unresolved transition array 
(UTA, Behar \et\ 2001)
indicated in the XSTAR model fits (fig. 10). The zero velocity wavelengths of the principal K-shell emission lines
and radiative recombination continua (RRC) falling in the 8--38 \AA\ waveband are indicated on the figure, and
several are clearly detected. 

To quantify the individual spectral features we modelled the RGS data with a simple power law, with
$\Gamma$$\sim$2.45 yielding a reasonably good continuum fit ($\chi^{2}$ = 4264/4084) over the 8--38 \AA\ band, and
then added gaussians to each candidate emission line, in turn, with wavelength, line width and flux as free
parameters. 4 emission lines were formally detected, those of OVIII Ly$\alpha$, the resonance (r) and forbidden (f) lines
of the OVII 1s-2p triplet, and the forbidden line of NVI, together with the RRC of OVII. Table 1 lists the
results. Interestingly, the gaussian line fits support the visible impression from the fluxed spectra that the
profiles of the resonance lines of  OVII and OVIII are resolved, the best fit line widths  corresponding to a
velocity width of 7000$\pm$3000 km s$^{-1}$. 

Figure 12a shows the OVII and VIII lines at a higher resolution (35 m\AA\ bin width), and is compared in figure
12b with the same  spectral region  observed in the low state spectrum. Due to the much poorer statistics the
latter fluxed spectrum is more coarsely binned, at 175 m\AA, and shows why only the OVIII Ly$\alpha$ and OVII (f)
lines were identified in the earlier analysis of the low state RGS spectrum  (Paper 1). The improved statistics of
the full spectrum now allows the OVII (r) line to be resolved,  though the total line fluxes remain consistent
(within a factor $\sim$1.5)
with those observed in the low flux state (Paper 1). That consistency is significant in the context of our
identification of a constant `core' soft emission (section 9), since the observed overall $\sim$0.3-1 keV flux -
including the variable continuum component - varies by a factor $\sim$4 over the 5 new \xmm\ observations (figure 1).

\section{Reconciling the RGS and EPIC spectra}

The most important finding from our analysis of the EPIC spectra is of a substantial column density of low
ionisation matter affecting the  strongly variable emission component. 

At first sight, the RGS data seem to be in conflict with the above picture, since no narrow absorption features
are observed, even though the EPIC ratio spectra (figure 9b) shows significant broad band absorption remaining up
to the mid-high flux state.  However, we note the RGS is designed to detect emission and absorption lines, and is
much less sensitive than EPIC to detecting continuum absorption.  It is also possible that the absorbing material,
if located close to the continuum emission region, is velocity broadened.  In that context we note that the OVII
and OVIII resonance emission lines are marginally resolved, suggesting an origin in moderately ionised matter with
a velocity width of 7000$\pm$3000 km s$^{-1}$, while it is conceivable that a still higher velocity/ more
turbulent outflow at smaller radii is responsible for the main absorption in the EPIC difference spectra. The width of
the Fe K line in the intermediate state spectrum (section 3) indicates such a higher velocity dispersion, if interpreted
as arising from fluorescence in the overlying absorber. 

To check
for comparable absorption edge structure in the RGS data we subtracted the low state data from the summed spectra for
March-June-September 2003, producing an intermediate state RGS difference spectrum, and plot that in Figure 13
against a simple power law model. With coarser binning (than in fig. 11)  the plot is consistent with absorption
edges similar to those found in the XSTAR fits to the intermediate state EPIC difference spectrum (fig. 6b),
though the statistics are not good enough to usefully constrain any velocity broadened edges.

\section{A re-appraisal of the Soft X-ray Excess}

Modelling the soft excess (conventionally defined as the excess soft X-ray flux below above an extrapolation of
the 2--10 keV power law) with one or more blackbodies, as in section 3, is a common practice in X-ray astronomy.
However, since the implied blackbody temperatures are much higher than appropriate to an AGN accretion disc, a
common explanation is that the disc photons gain energy by electron scattering in a hotter `skin' or a `corona'
lying above the disc. Nevertheless, such Comptonisation models have remained rather ad hoc. Recently, a scaling by black
hole mass of (more robust) Comptonisation models for Galactic black hole sources (Done and Gierlinski 2003) failed
to explain the sharp upturn often seen below $\sim$1 keV in AGN  spectra. Noting also the similar shape (or
blackbody temperature) of soft excess in AGN over a wide luminosity range, those authors proposed  that the soft
excess in AGN could be an artefact of `unseen or ignored' absorption (Gierlinski and Done, 2004). 

The idea that absorption could be playing a larger part than normally assumed in shaping the broad band spectra of
Seyfert 1 galaxies was put forward in an early mini-survey of \xmm\ observations by Pounds and Reeves (2002).  In
figure 1 of Pounds and Reeves (2002) we showed the similarity in the `observed' broad band spectra of AGN over a
wide luminosity range, while noting that  the more luminous sources exhibited a more `gradual' onset of the soft
excess. In comparing the present EPIC spectra of \1h\ we now find the same qualitative trend with flux level.
Figure 14 compares the apparent soft excess above a 2--10 keV power law for the 5 flux states of \1h, with the
higher flux states showing a more `gradual' (or hotter) soft excess. 

Our present analysis suggests that the `conventional' soft excess is indeed strongly affected by absorption. The
additional point to emphasise here is that the individual difference spectra, for high, intermediate, and mid-low
states of \1h\ exhibit {\it no soft excess}. Instead, we identify a {\it core soft X-ray emission} component in the
extreme low flux observation of September 2002 (Paper 1). Thermal emission from the accretion disc modified by
scattering in a hotter skin or corona and reflection from a hot inner disc surface are possible contributors to
this `core' soft component. However, our present analysis of \1h, with evidence for a substantial column of cold
nuclear gas becoming less opaque as the X-ray flux increases, suggests that recombination emission from associated
photoionised (outflowing?) matter is a natural origin of the soft excess. To test that idea we can compare the
`core' soft excess observed in the \xmm\ low state observation of \1h\ with an emission model grid from XSTAR.

As with the absorption grids, the ionising continuum is a power law of energy index -1. The key variable parameters
in this test were the ionisation state and relative metal abundances, which essentially determine the
emission spectrum for comparison with the overall shape of low state soft excess observed in EPIC, and with
the (few) strongest features identified in the RGS spectrum.  

Figure 15a shows the `soft X-ray emission component', illustrated by removal of the black body and gaussian emission line
attributed to a blend of the OVII triplet and OVIII Ly$\alpha$ from the broad band fit to the low flux state EPIC
spectrum discussed in Paper 1. Quantitatively, the X-ray flux in that soft excess (0.3--1 keV) is
$\sim$1.3$\times$10$^{-12}$ ergs s$^{-1}$, some 30 percent of the flux removed by the low energy absorption trough
in the intermediate flux state (`mean' ?) spectrum of \1h. Replacing the blackbody and O-K emission components
with a grid of photoionised emission models in XSTAR produced a statistically good fit ($\chi^{2}$=1028/1058 dof),
ie was able to reproduce the {\it shape} of the core soft emission, with an ionisation parameter log$\xi$$\sim$1.3 and
element abundances for C, N, O, Ne and Fe (relative to solar) of 0.4,0.4,0.15,0.35 and 0.45. We note the ratio of
the key elements, O and Fe, suggests an over-abundance of Fe, similar to that found in the absorption modelling.
Figure 15b  reproduces the fitted XSTAR emission  spectrum, showing how the emission profile, attenuated by the
Galactic absorption in the line of sight to \1h, matches the strongly peaked soft excess in figure 15a. It may
also be understood, qualitatively, how the only lines detected in the full RGS spectrum of \1h, velocity broadened
in the figure by a gaussian width of $\sigma$$\sim$5eV, are the principal emission lines  of OVII, OVIII, NVI and
CVI.

Finally, it is interesting to compare the Fe K line in the model of figure 15b with the emission line detected in
the intermediate flux state EPIC data (figures 2a and 3). Relative to the resonance lines of OVII and VIII, the
XSTAR model yields an Fe line which is a factor $\sim$10 weaker, and centered at $\sim$6.42 keV (corresponding
to FeXVII) compared with the EPIC line energy of 6.25$\pm$0.12 keV. It seems probable, therefore, that the bulk of
the observed Fe K line arises by fluorescence in low ionisation matter, with the absorbing column being an obvious
candidate. In that case the observed line width and mean energy both indicate the re-processing matter lies within
$\sim$100 Schwarzschild radii of the SMBH, while the equivalent width requires a substantial covering factor.

\section{Discussion}

The above analysis assumes that the underlying hard power law and `soft excess' seen in the low flux state of \1h\ in
September 2002 remains constant - or, at least, much less strongly variable - over a timescale of 1--3 years. The
similar power law index/ compTT parameters and absence of a soft excess in the mid-low, intermediate and high flux
state difference spectra support that assumption and - in turn - suggest the `core' and variable X-ray
emission components have a separate origin. 

Our good fortune in observing \1h\ over a wide range of flux levels has yielded three main results. First,
examination of the difference spectra of the intermediate and mid-low state EPIC data supports the conclusion in
Paper 1 that the variable emission component has the form of a steep power law, or cool Comptonised thermal
spectrum.  Second, this variable emission continuum shows the imprint at intermediate flux levels of absorption in
low ionisation matter, the effect of which decreases as the source flux increases. Third, we note that the X-ray
luminosity `lost' in the time-averaged absorbed flux is of the same order as that in the soft X-ray emission
component seen in the low state spectrum of \1h, reported in Paper 1, suggesting a possible origin of the `core'
soft X-ray emission.

Thermal Comptonisation remains a likely mechanism for the variable continuum emission component, with an
excellent match to the high state difference spectrum.  We also have evidence from simultaneous measurements of
\1h\ with the Optical Monitor on \xmm\ (Mason \et\ 2001) that the UV (2120, 2910\AA) flux  increased by $\sim$20
percent from the low to the intermediate flux state (with smaller increases of $\sim$10 percent in the U band and
$\leq$5 percent in V), consistent with the larger change in the UV between the high and low flux states (Paper 1).
These data appear to confirm a link between the X-ray spectral change and enhanced thermal emission from the
accretion disc. However, while increased disc emission could drive that change, the similarity in the luminosity
increase in the soft X-ray and UV bands leaves open the alternative possibility that the UV flux increase is due
to reprocessing of soft X-rays directed towards the (unobscured) outer disc. In the latter case the brightening
X-ray continuum would primarily be a consequence of an increased optical depth of Comptonising electrons. We note,
in passing, that if the scenario we propose for \1h\ is widely applicable to radio quiet AGN, then the hard
(disc-corona emission?) and
softer continuum components must be coupled in some way in order to explain the `canonical' power law index
(neglecting reflection) of $\Gamma$$\sim$1.7 at 2--10 keV, observed in broad-band fits for Seyfert 1 galaxies in
general (Nandra and Pounds 1994), as well as the long-term average spectrum of \1h.

Given that the peak luminosity of the variable X-ray emission component is $\leq$10 percent of the bolometric
(accretion) luminosity of \1h, the kinetic energy in a wind could be sufficient to support a separate emission process.
Although we have found no unambiguous evidence for an outflow in \1h, the resolved emission  line profiles of Fe
K, OVII and OVIII imply a large velocity dispersion, while it seems likely that as the cold absorber is
photoionised some fraction will be driven away by the strong continuum radiation pressure.  We also note the
growing evidence in other recent studies for ionised outflows at velocities of $\sim$5--20 percent of c (Chartas
\et\ 2002; Pounds \et\ 2003, 2004b; Chartas \et\ 2003; Reeves \et\ 2003). Any such outflow is likely to have an
initial velocity which reflects the gravity of its origin (as with stellar winds), so that a wind emerging from a
small radius will have a correspondingly high velocity.  The kinetic energy carried by such a `black hole wind'
(King and Pounds 2003) could then support a separate emission mechanism of order v/c times the bolometric
luminosity, with shocks in the outflow (eg. Camenzind and Courvoisier 1983) perhaps providing the enhanced Comptonising 
electrons.  A timescale for major spectral change in \1h\ being of order several months, one possible
cause of an increased outflow may be the co-alignment of magnetic field lines in the inner disc (Livio \et\ 2001).
Alternatively, if the variable X-ray emission arises from the base of a jet (eg. Markoff \et\ 2001), then it is
interesting to note that synchrotron radiation could - qualitatively - also explain the spectral form of \1h.

Absorption features in both the difference and flux-ratio spectra of \1h\ provide clear evidence for substantial
low ionisation matter overlying - or embedded in - the X-ray continuum emission region. The marked decrease in
opacity and/or covering factor of the absorber, as the X-ray flux increases, indicates the absorbing matter lies
close to the SMBH, a conclusion also consistent with the Seyfert 1 classification of \1h.  The low ionisation
parameter then requires the absorbing matter to be of high density ($\ga$$10^{17}$ cm$^{-3}$) to survive the
intense  continuum irradiation. A possible scenario is for the absorbing matter to be in small, dense clouds which
partially cover the continuum source (eg. Ferland and Rees, 1988), the covering factor decreasing as the source
flux increases, a change modelled by a reducing column density in our spectral fitting. Dense matter at the base
of a jet or outflow provide other potential sites. The recent discussion of `aborted jets', where dense blobs of matter
are ejected from the inner accretion disc at sub-escape velocities (Ghisellini \et\ 2004), also offers a geometry
which might be compatible with the results we report here.

Although the spectral variability in \1h\ is extreme, the low flux state spectrum is remarkably similar  in
appearance to the low state spectra of two other bright Seyfert 1 galaxies, NGC4051  (Pounds \et\ 2004b, Uttley
\et\ 2004), and MCG-6-30-15 (eg Reynolds \et\ 2004).  In each case a hard power law continuum also exhibits a
broad spectral feature at $\sim$3--6 keV that may alternatively be fitted by an extreme relativistic Fe K line or
by partial covering of the X-ray continuum by low ionisation matter. In the disc-corona model of hard X-ray emission,
where UV photons from the accretion disc are up-scattered in an overlying corona (eg. Haardt and Maraschi, 1991),
an unusually hard spectrum indicates a relatively low UV flux and `photon-starved' corona.  The intrinsic hardness
of the `core' spectrum would be reduced somewhat if continuum reflection is enhanced, for example by  light
bending in the  strong gravity near the black hole, as proposed by Miniutti and Fabian (2004) to explain the
extreme Fe K emission line in the low state spectrum of MCG-6-30-15. In that case a simultaneous observation by
\sax\ supported the enhanced reflection, rather than partial covering models. However, in the light of the new
evidence reported here for substantial cold matter close to the  black hole in \1h, it is hard to exclude the
possibility that absorption also modifies the strength and profile of the key diagnostic Fe K emission line in AGN
spectra.

The strong forbidden line of OVII is evidence that a substantial part of the `core' soft X-ray emission comes from
an ionised gas of relatively low density, where the emission measure then yields a radial extent (Paper 1) sufficient to
maintain the soft X-ray emission essentially constant  over the 3 years of our \xmm\ studies of \1h. The detection
of resolved resonance lines of OVII and OVIII, and a strong RRC of OVII in the full RGS spectrum suggests an 
additional component to the soft X-ray emission from higher density, and perhaps turbulent, recombining gas closer
to the continuum source. The comparable luminosity of the `core' soft X-ray emission with that of the absorbed
continuum in the intermediate (average) flux state then offers a natural explanation of the soft X-ray emission, as
an ionised outflow subsequently recombines. The `true' soft excess in Seyfert 1 galaxies, as we find for \1h, may then
be more akin to the soft X-ray emission seen in Seyfert 2 galaxies, and arising from an extended region of ionised
gas. 

Finally, we point out the detection of cold absorbing matter close to the SMBH in a luminous Seyfert 1 galaxy
implies an additional component to the `standard model' by which type 1 and type 2 AGN are distinguished in
relation to absorption in a distant torus (Antonucci 1993).

\section{Summary}

1.  A series of \xmm\ observations of the luminous Seyfert 1 galaxy \1h\ has shown the large-scale spectral
variability is primarily due to a steep power law or cool Comptonised thermal emission component. In seeking an
alternative to the disc-corona model for this new emission component it is interesting to note that mechanical
energy in a wind could be sufficient to support such a process, perhaps via shocks in a turbulent or inhomogeneous
outflow.

2.  Broad absorption features superimposed on the variable emission continuum require substantial cold, dense
matter apparently lying close to the central SMBH. As \1h\  brightens from a low to a high flux state the
ionisation parameter of this absorbing matter increases and its column density (or covering factor)  falls. 

3.  The existence of cold absorbing matter close to the SMBH in a luminous Seyfert 1 galaxy adds a new component
to the standard model in which type 1 and type 2 AGN are distinguished in terms of an obscuring torus. A primary
difference is that this inner absorber is subject to the intense radiation - and perhaps coherent magnetic fields
- which will control its opacity and probably drive it outward in the form of a wind. 

3.  An underlying assumption in our analysis is that a hard `core' spectral component remains essentially
unchanged over the  1--3 year period of the observations. We note the similarity of this `core' X-ray spectrum of
\1h\ to high quality `low state spectra' of other Seyfert 1 galaxies, including NGC 4051 and MCG-6-30-15, which
also show a flat power law, relativistic Fe K emission line or partial covering, and strong `soft excess'.

4.  In the case of \1h\ we find the variable soft excess is essentially an artefact of continuum absorption, while
a `core'  `soft X-ray emission' component has a spectral form and luminosity consistent with re-emission of the
absorbed X-ray continuum in an extended region of ionised gas. 

5.  While the X-ray spectral variability of \1h\ is extreme, its explanation in terms of a variable soft emission
component, modified by absorption in low ionisation matter close to the SMBH, is unlikely to be unique, suggesting
the need for re-appraisal of other high quality AGN X-ray spectra. 

6.  Finally, we note the occurrence of a substantial column density of cold gas overlying the
hard X-ray source in a luminous AGN has implications both for the diagnostic potential of the Fe K emission
line and the new source population required to explain the hard ($\ga$5 keV) Cosmic X-ray Background spectrum.

\section{Acknowledgments}

The results reported here are based on observations obtained with \xmm, an ESA science mission with instruments
and contributions directly funded by ESA Member States and the USA (NASA). The authors wish to thank the SOC and
SSC teams for organising the \xmm\ observations  and initial data reduction, Tim Kallman for provision of a new
XSTAR grid, and the referee for constructive comments on the initial text. KAP is pleased to acknowledge a
Leverhulme Trust Emeritus Fellowship.

\clearpage

\begin{figure}
\rotatebox{-90}{
\epsscale{0.7}
\plotone{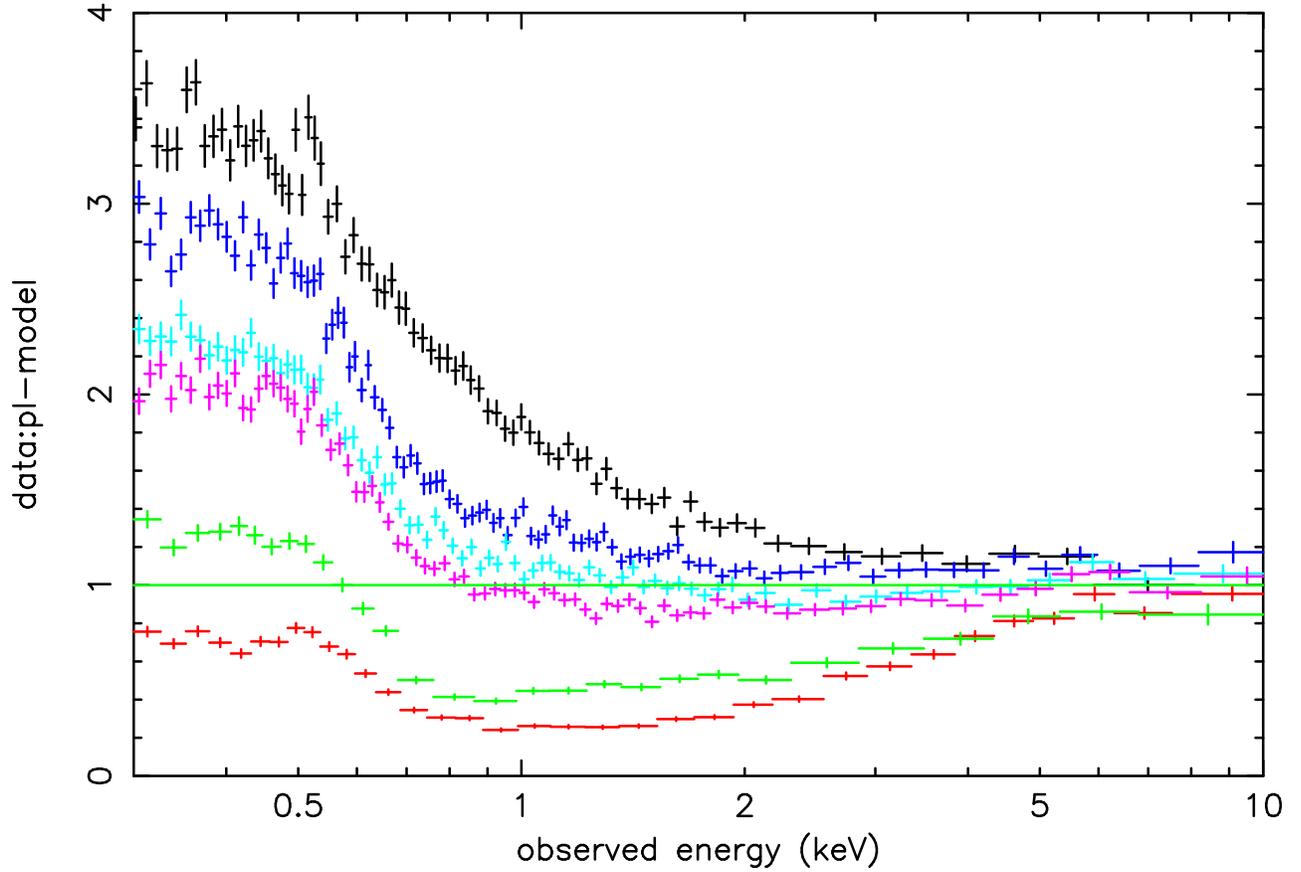}}
\caption{EPIC spectral data from the observations of 
December 2000, orbit 181 (black); September 2002, orbit 512 (red); December 2002, orbit 558 (green); 
March 2003, orbit 605 (dark blue); June 2003, orbit 649 (light blue); and September 2003, orbit 690 (magenta); 
compared (at 2--10 keV) with a power law
of photon index $\Gamma$=1.73. 
For clarity only the pn camera data are shown.  \label{fig1}}
\end{figure}

\clearpage

\begin{figure}
\rotatebox{-90}{
\epsscale{0.3718}
\plotone{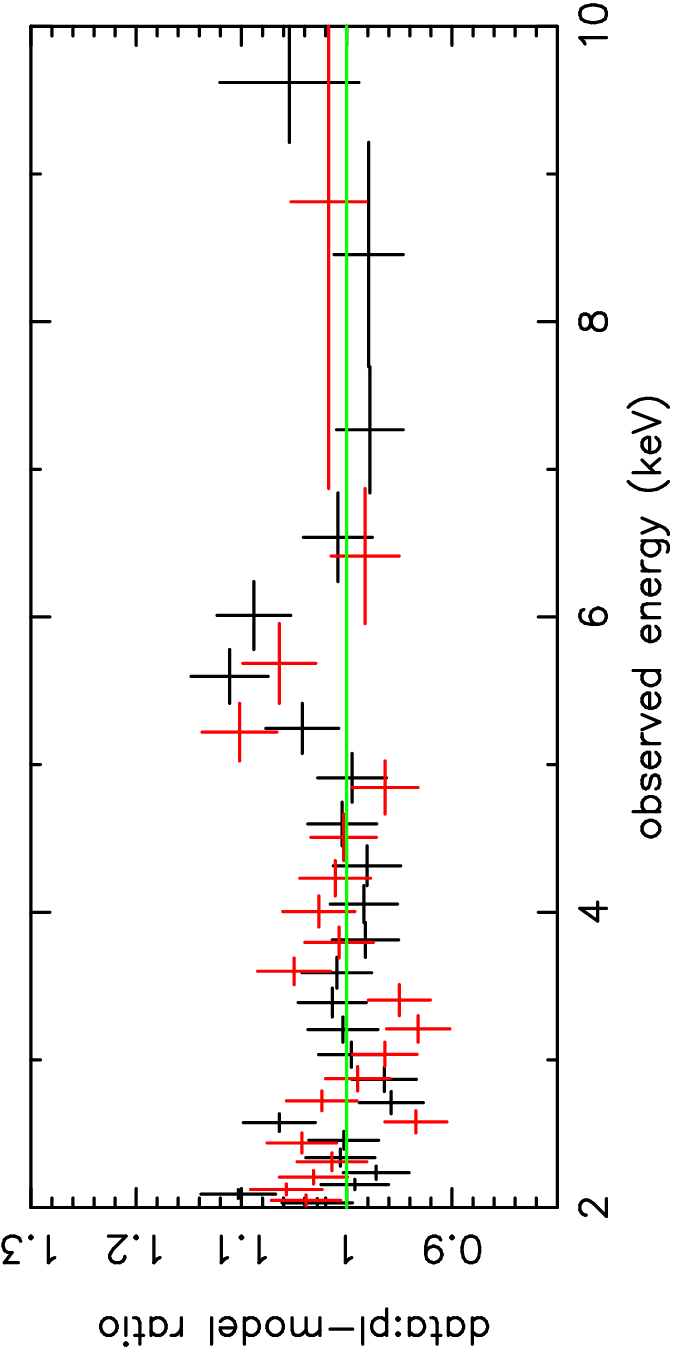}}
\rotatebox{-90}{
\epsscale{0.358}
\plotone{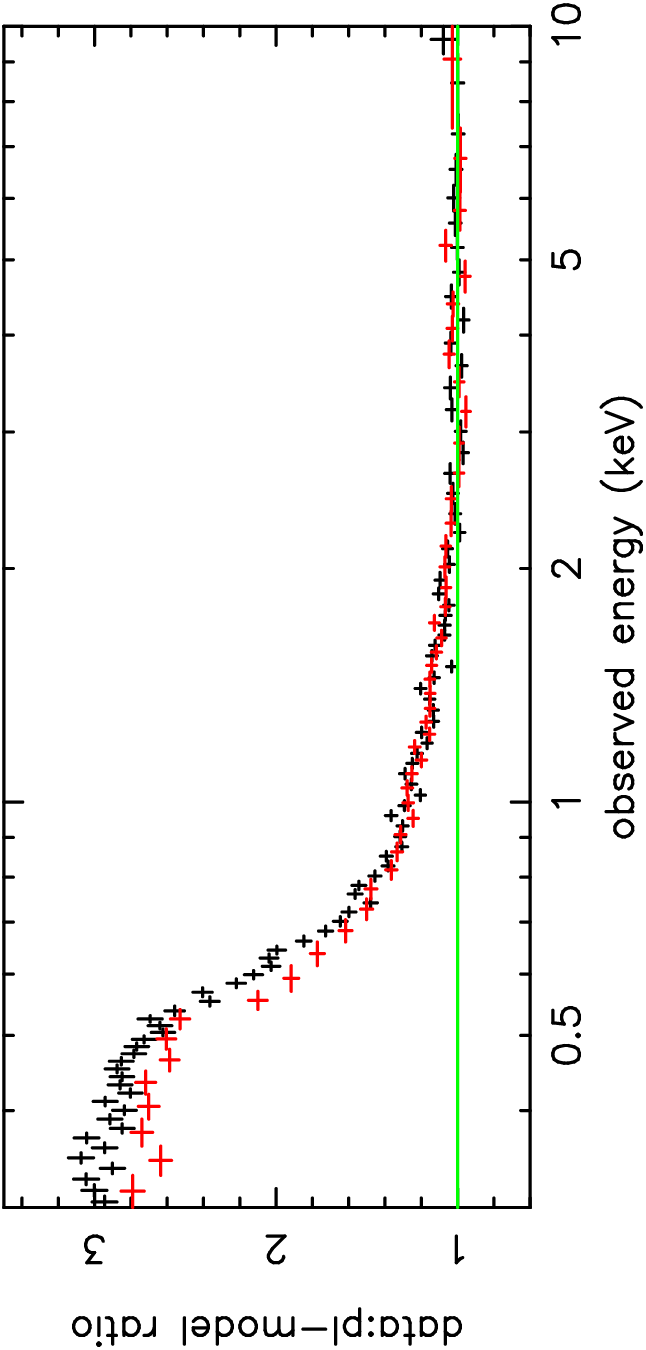}}
\rotatebox{-90}{
\epsscale{0.34}
\plotone{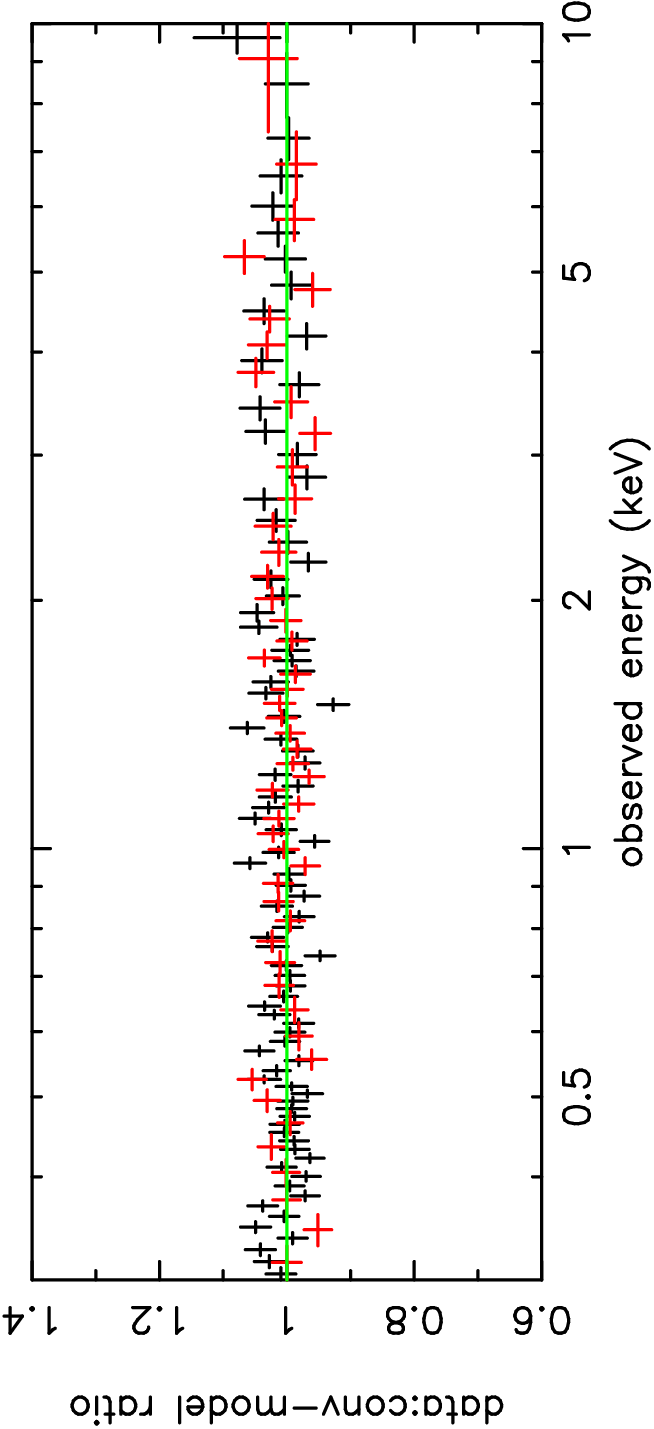}}
\caption{a.(top) Ratio of pn (black) and MOS data (red) to 2--10 keV power law fits to the intermediate flux state data
(June/September 2003), showing a weak Fe K emission line. b.(mid) Extrapolation of 2--10 keV power law to
0.3 keV showing a strong
soft excess. c.(lower) Ratio of data to conventional multi-component model spectrum described in section 3. 
\label{fig2}}
\end{figure}

\clearpage

\begin{figure}
\rotatebox{-90}{
\epsscale{0.7}
\plotone{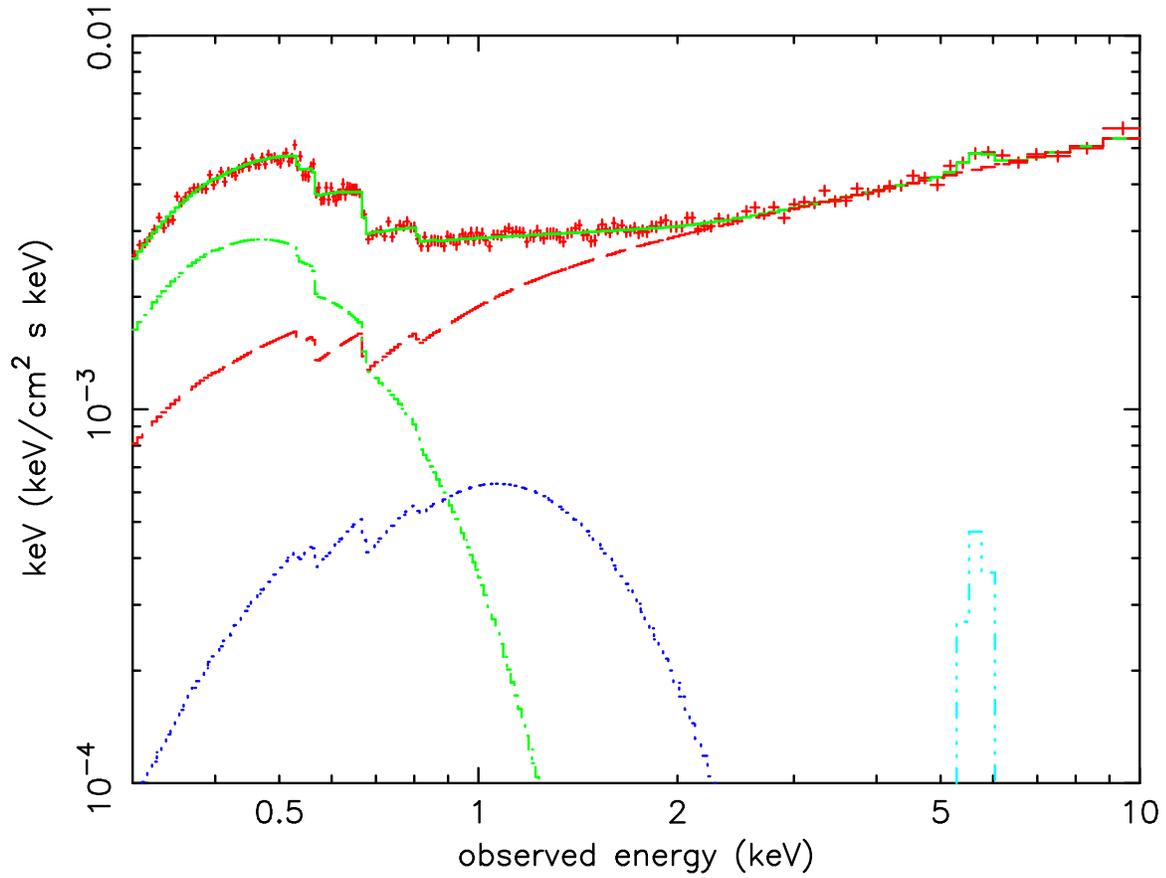}}
\caption{Unfolded broad band spectrum fitted to the intermediate flux state data for \1h. The spectral components 
of this conventional fit are: power law (red), blackbodies (dark blue and green) and Fe K emission line (light blue). For clarity only the 
pn data are shown. \label{fig3}}
\end{figure}

\clearpage

\begin{figure}
\rotatebox{-90}{
\epsscale{0.32}
\plotone{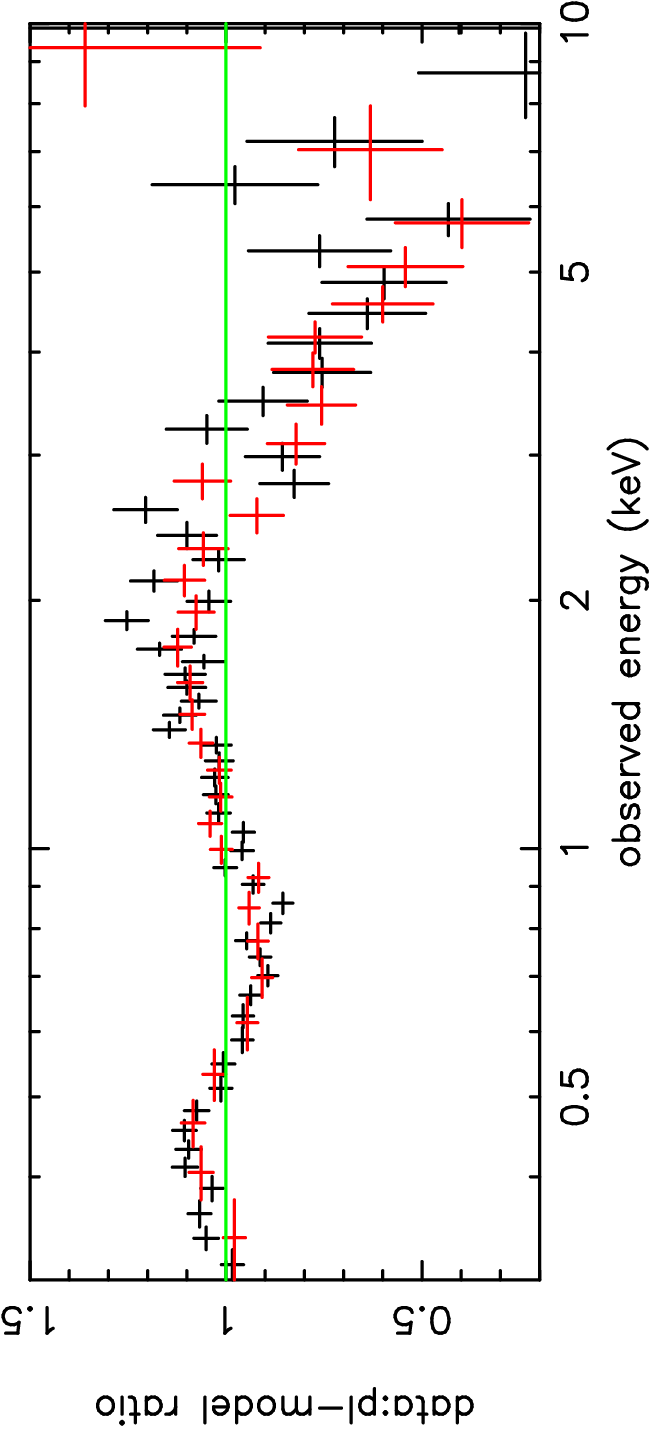}}
\rotatebox{-90}{
\epsscale{0.307}
\plotone{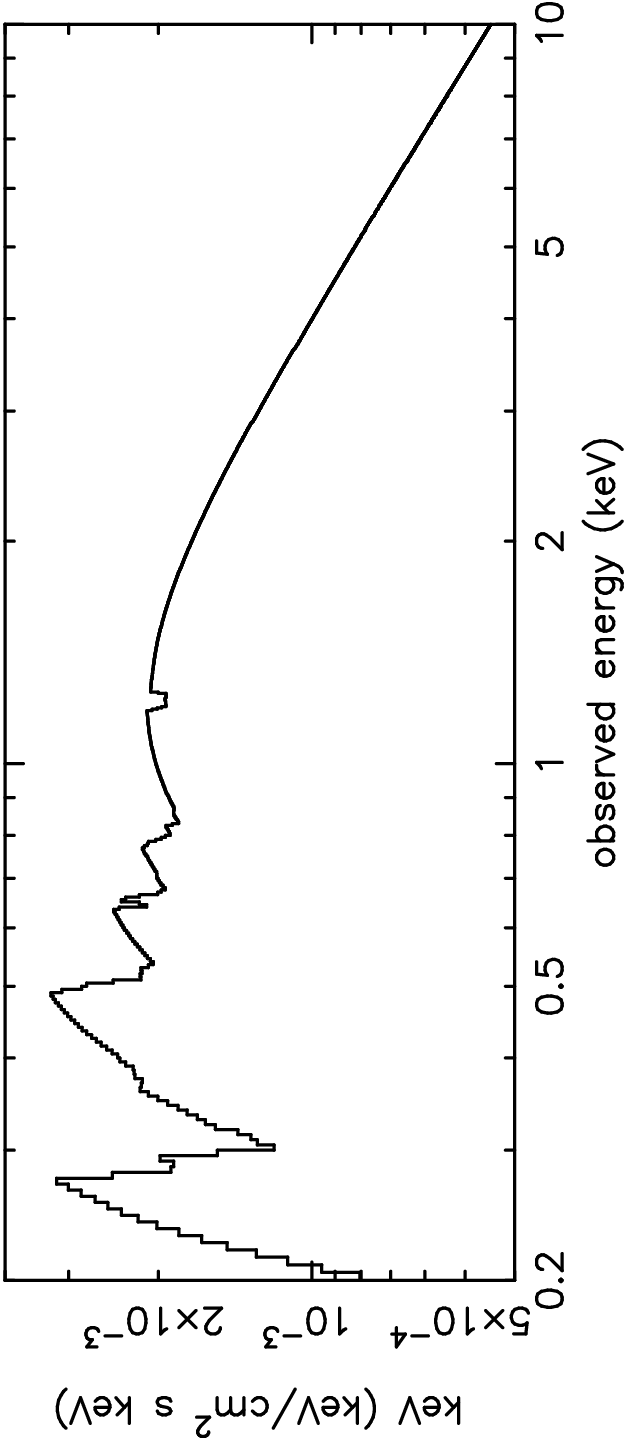}}
\rotatebox{-90}{
\epsscale{0.32}
\plotone{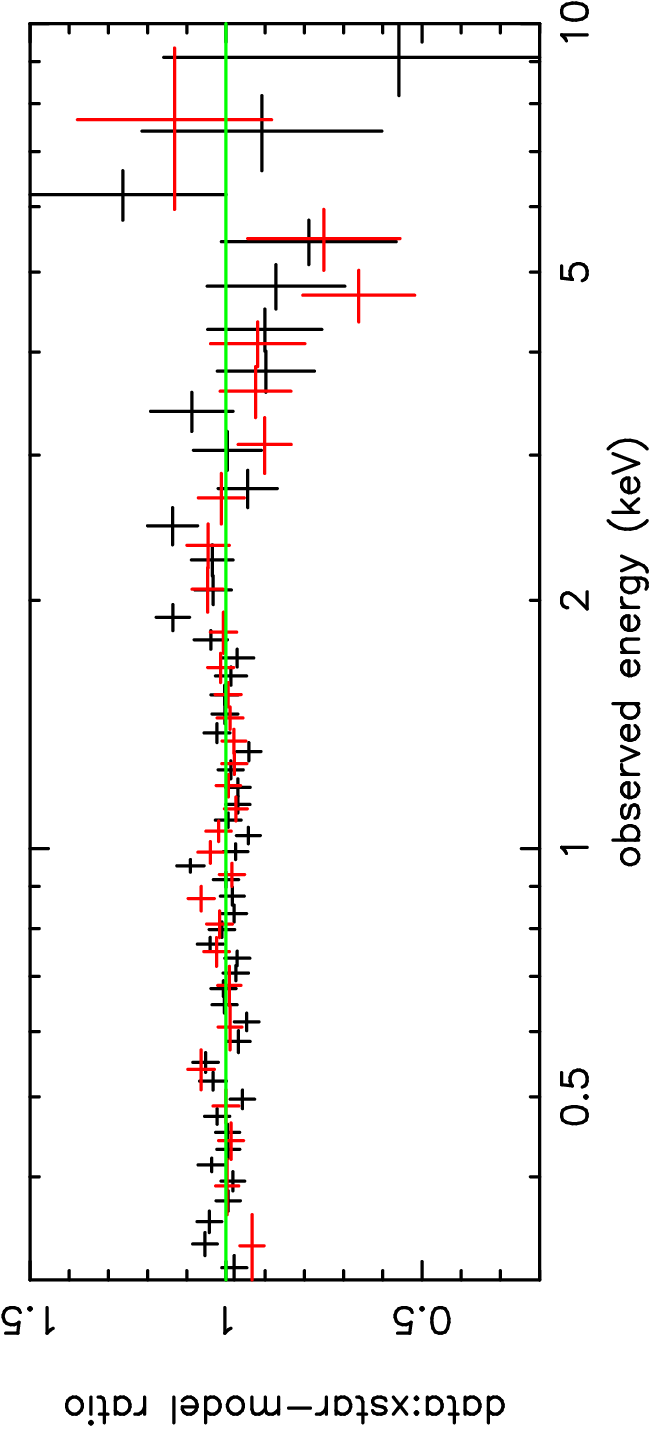}}
\caption{a.(top)Ratio of data to a single power law fit for the intermediate state difference spectrum. 
b.(middle)Power law plus XSTAR absorption model for the same difference spectrum, with strong absorption edge 
structure (in order
of increasing energy) of C, O, Fe and Ne.  c.(lower) Ratio of difference spectrum data to 
the power law plus absorption modelled in XSTAR. See section 4 for details. The small spectral feature seen close to the Mg
edge at $\sim$1.2 keV is due to curtailment of the tabulated photoionisation cross sections.\label{fig4}}
\end{figure}

\clearpage

\begin{figure}
\rotatebox{-90}{
\epsscale{0.35}
\plotone{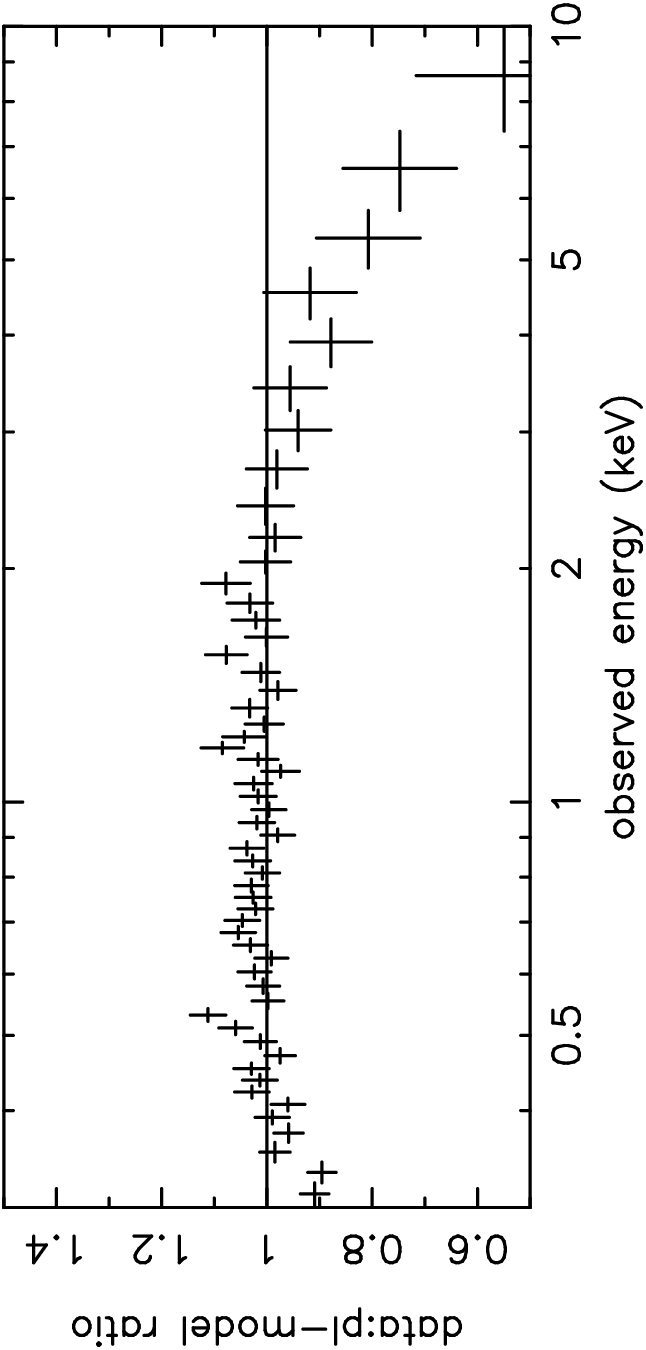}}
\rotatebox{-90}{
\epsscale{0.35}
\plotone{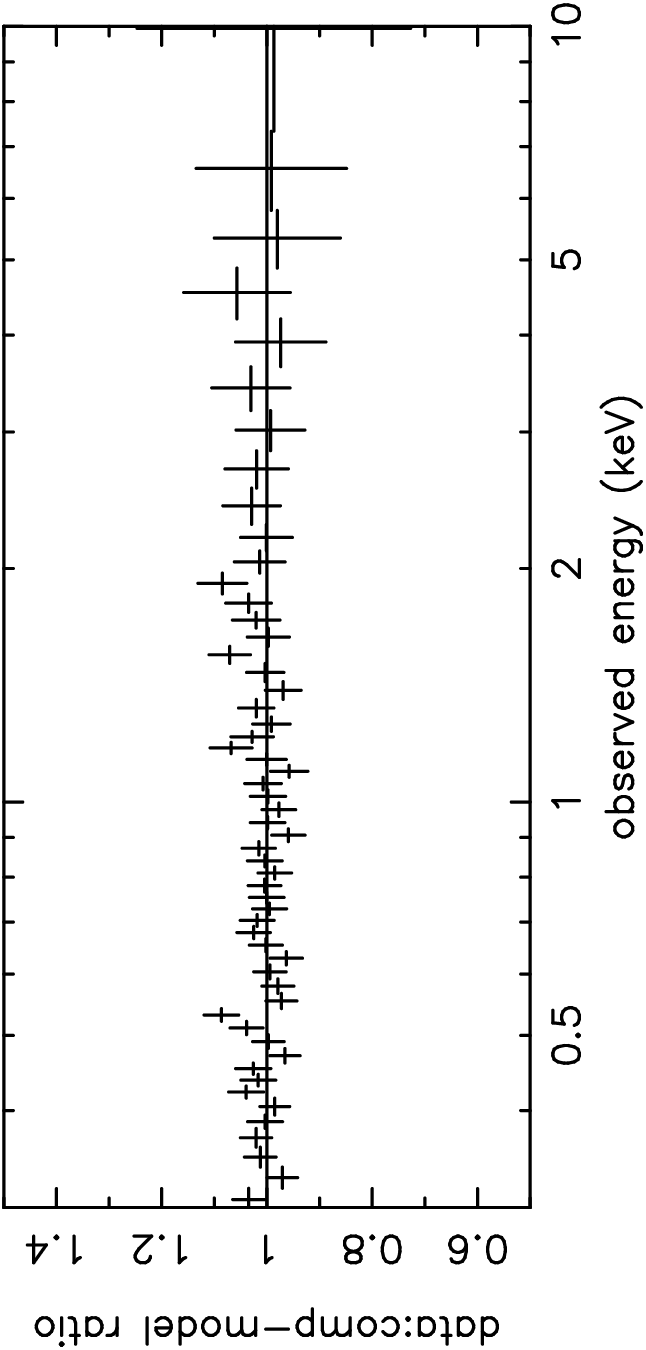}}
\caption{a.(top) Ratio of high state difference spectrum (December 2000) to a single power law model of photon 
index 2.47. b.(lower) Ratio of same data
to a thermal Comptonisation model, with seed photons of kT=73$\pm$3 eV scattered in a plasma of
temperature kT=2.7$\pm$0.6 keV and optical depth $\tau$=4.4$\pm$0.6. \label{fig5}}
\end{figure}

\clearpage

\begin{figure}
\rotatebox{-90}{
\epsscale{0.34}
\plotone{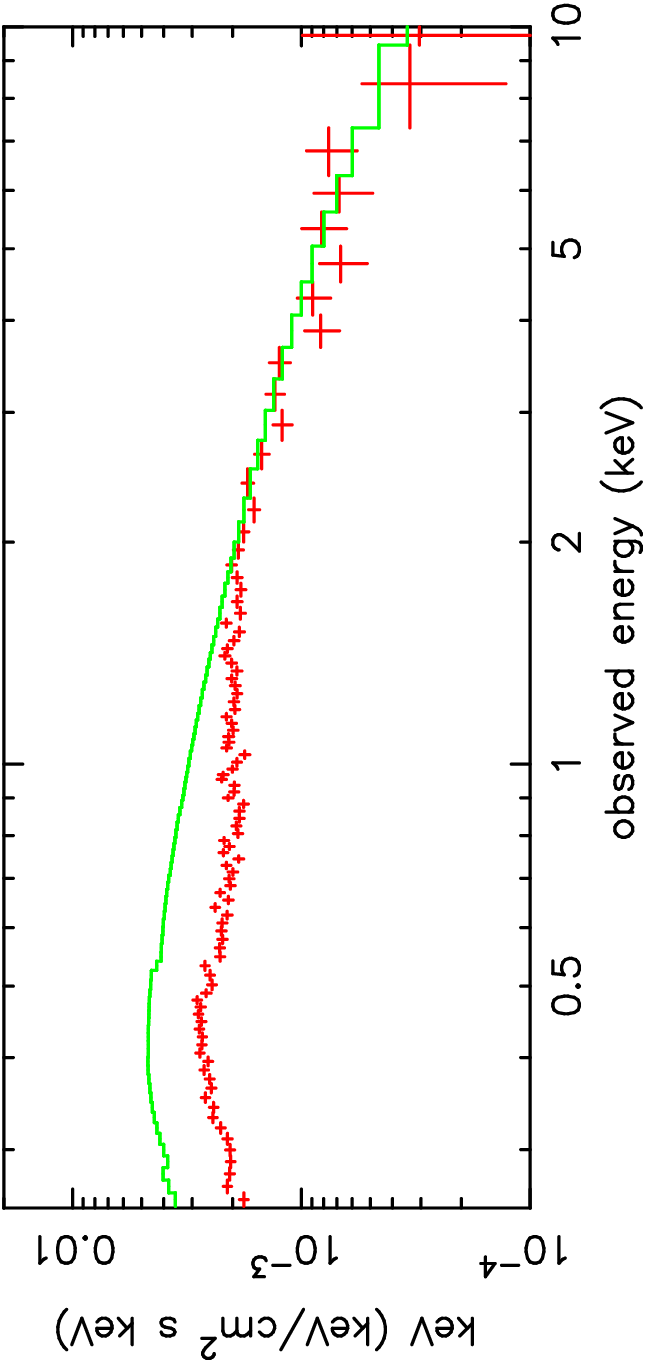}}
\rotatebox{-90}{
\epsscale{0.34}
\plotone{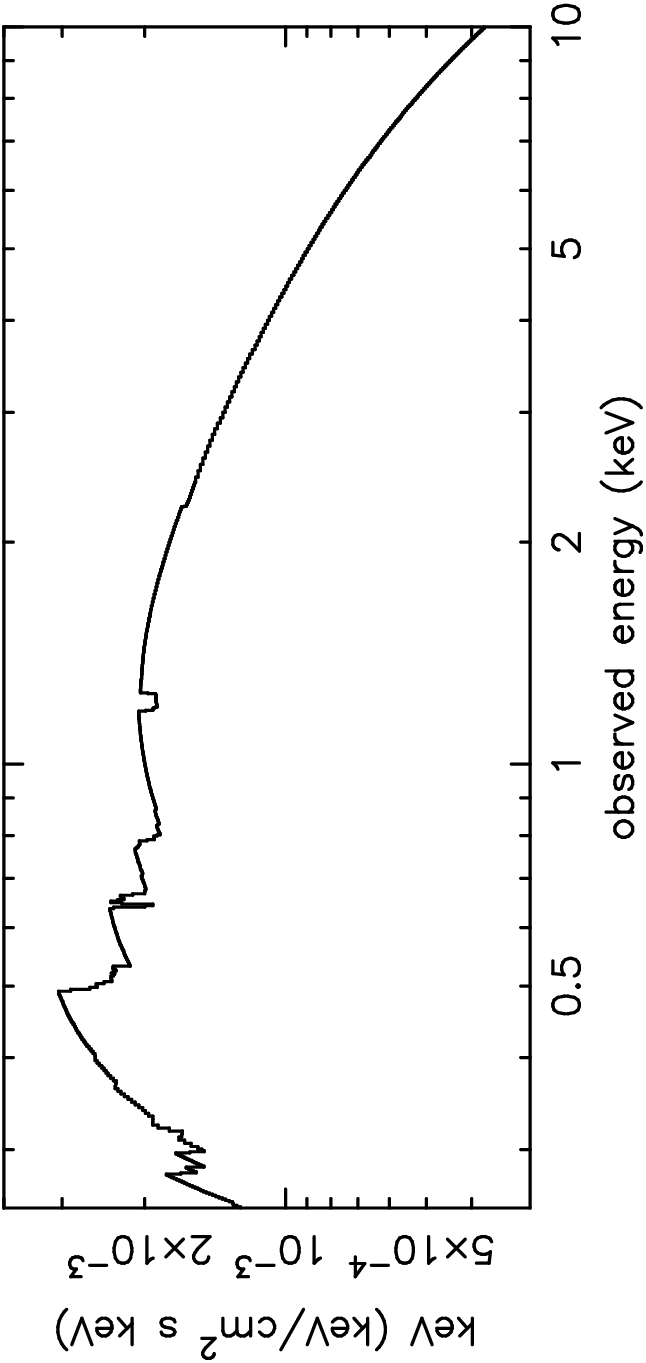}}
\rotatebox{-90}{
\epsscale{0.32}
\plotone{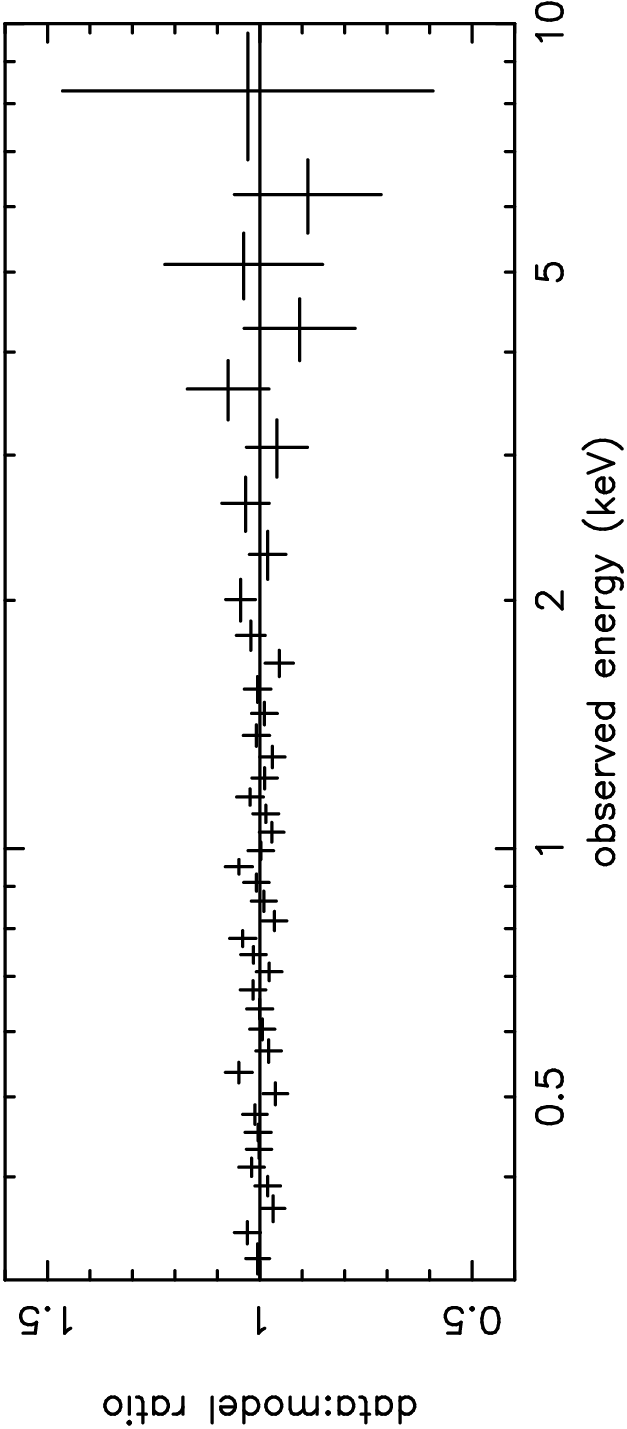}}
\label{fig6}
\caption{a.(top) Comptonised emission component and pn camera data for the intermediate state difference spectrum 
of \1h. 
b.(mid) Comptonised emission plus XSTAR absorption model for the same data. c.(lower) Data to model ratio.
\label{fig6}}
\end{figure}

\clearpage

\begin{figure}
\rotatebox{-90}{
\epsscale{0.35}
\plotone{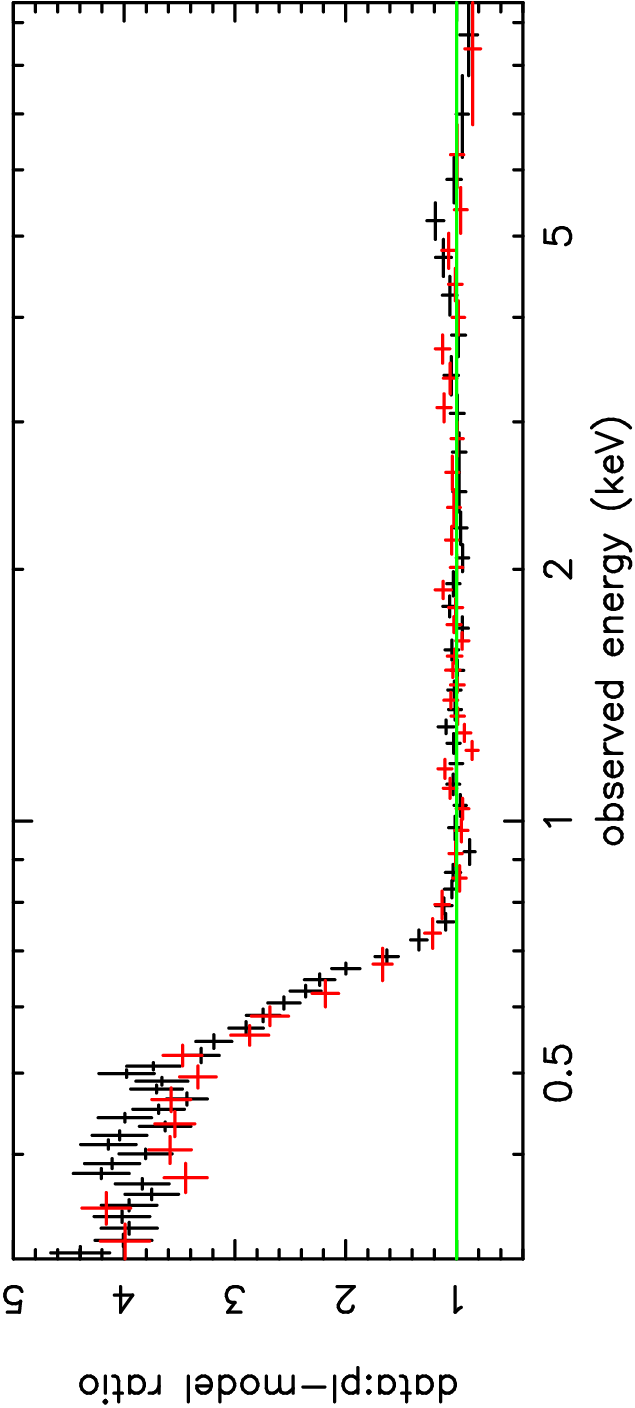}}
\rotatebox{-90}{
\epsscale{0.35}
\plotone{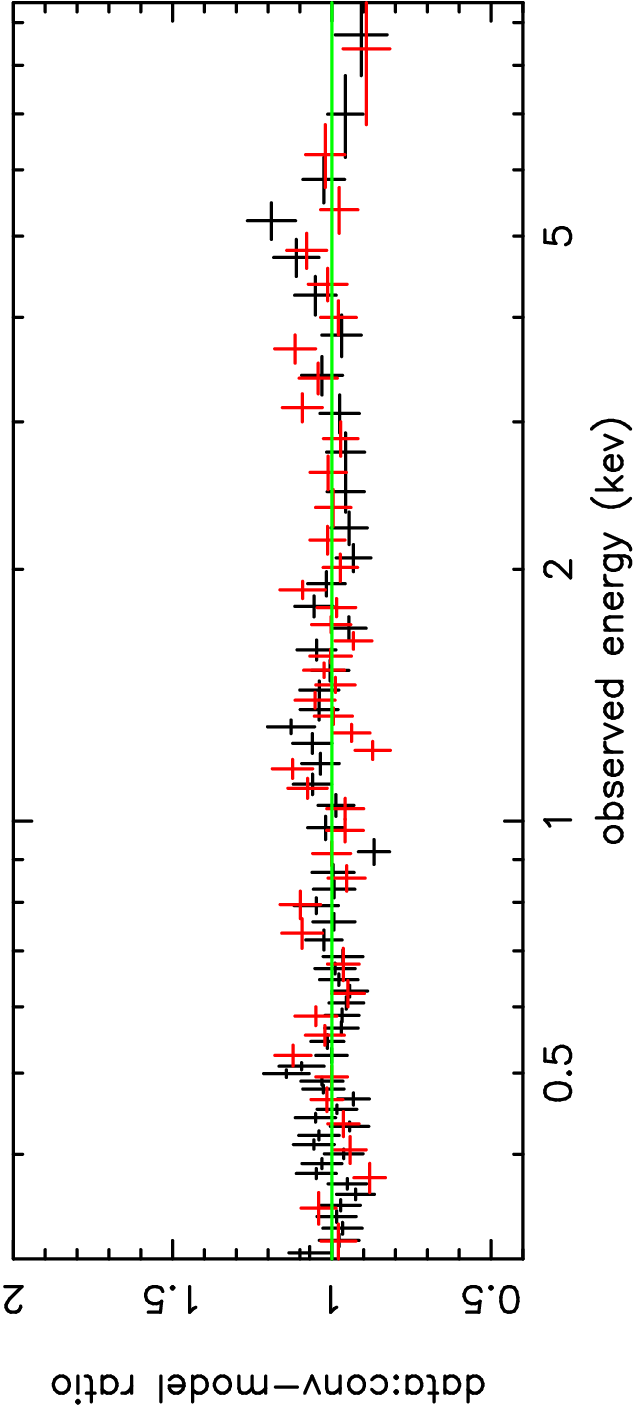}}
\caption{a.(top) Ratio of pn (black) and MOS data (red) to 2--10 keV power law fits to the mid-low flux state data
(December 2002). b.(lower) Ratio of data to conventional multi-component model spectrum described in section 5. 
\label{fig7}}
\end{figure}

\clearpage

\begin{figure}
\rotatebox{-90}{
\epsscale{0.33}
\plotone{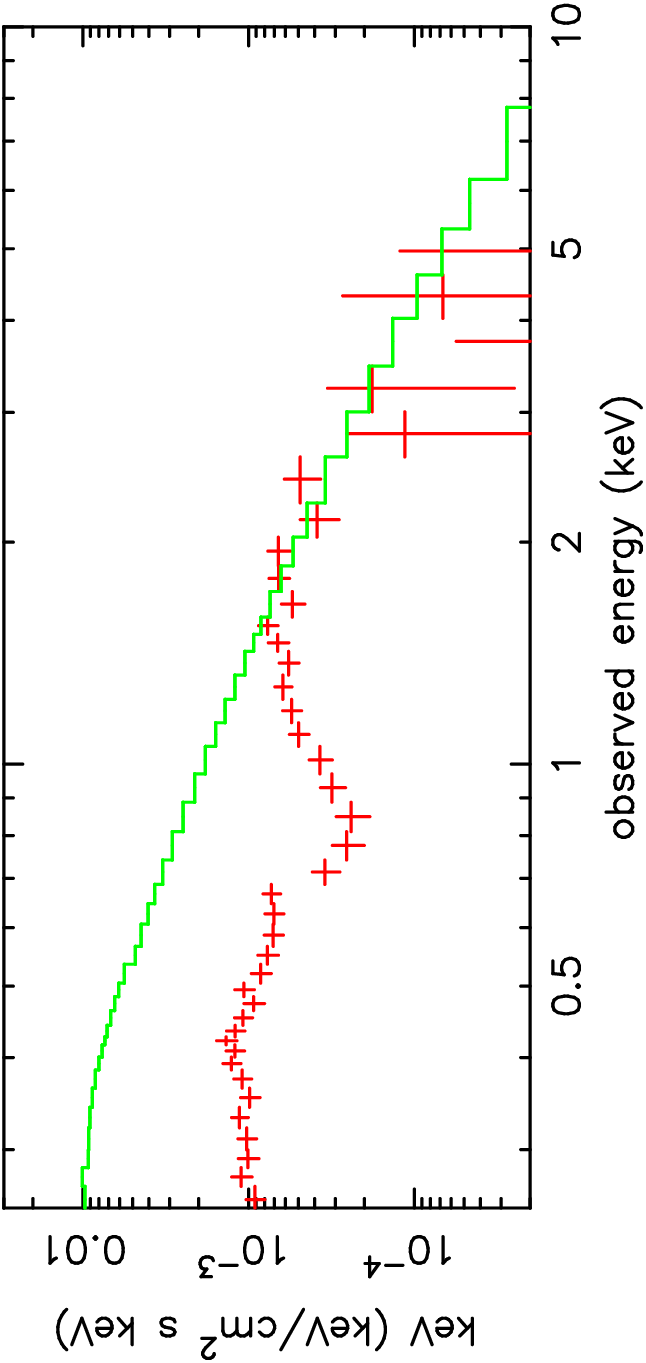}}
\rotatebox{-90}{
\epsscale{0.33}
\plotone{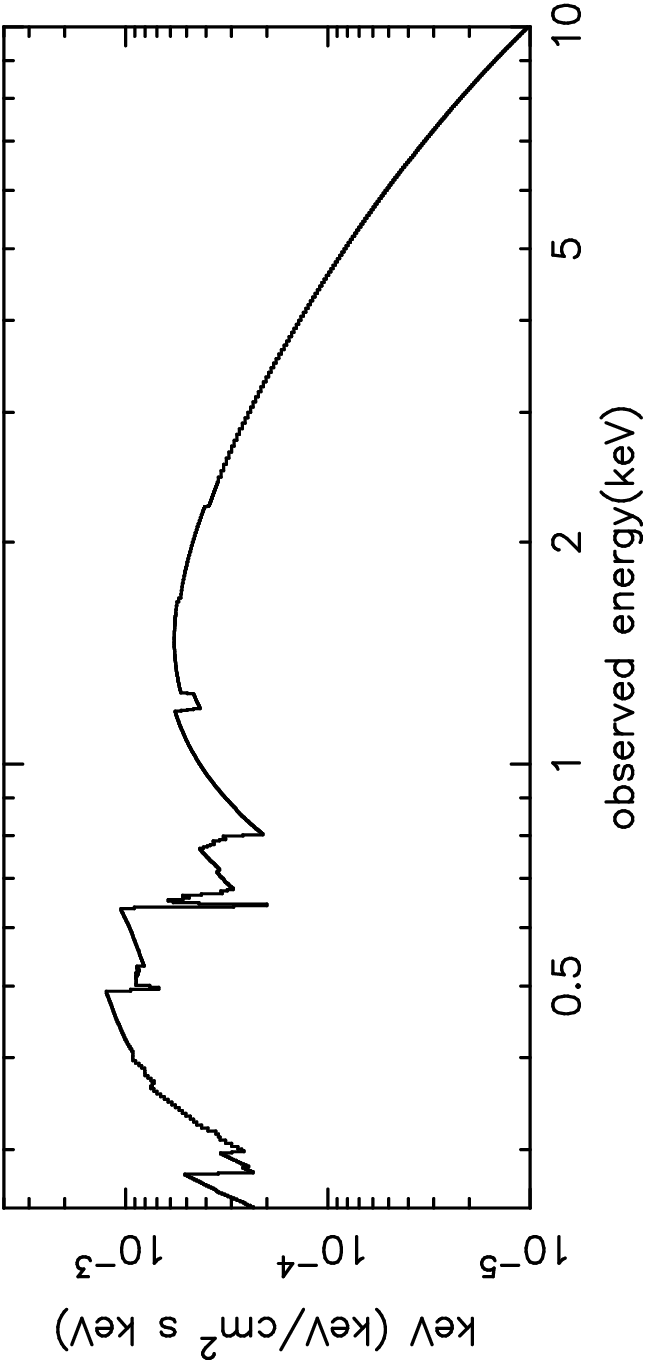}}
\rotatebox{-90}{
\epsscale{0.33}
\plotone{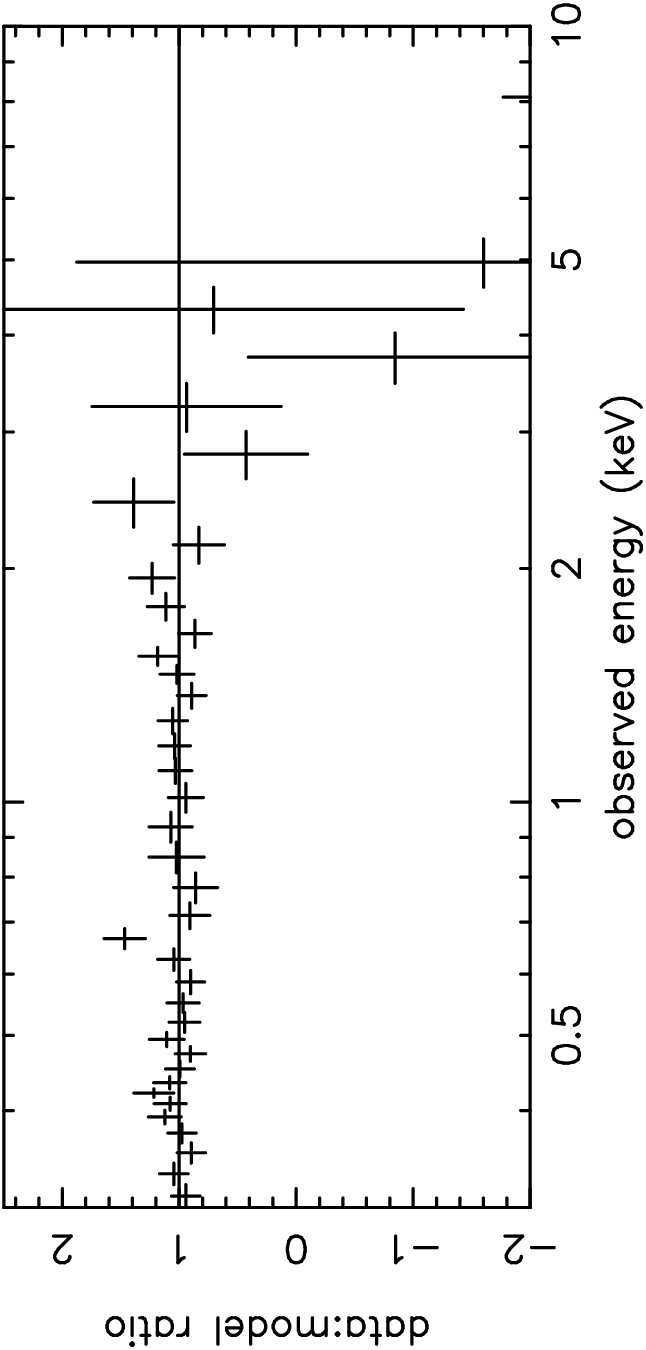}}
\caption{a.(top) Comptonised emission component and pn camera data for the mid-low state difference spectrum of 
\1h. b.(mid) 
Comptonised emission plus XSTAR absorption model for the mid-low state difference spectrum. c.(lower) Data to 
model ratio. 
\label{fig8}}
\end{figure}

\clearpage

\begin{figure}
\rotatebox{-90}{
\epsscale{0.37}
\plotone{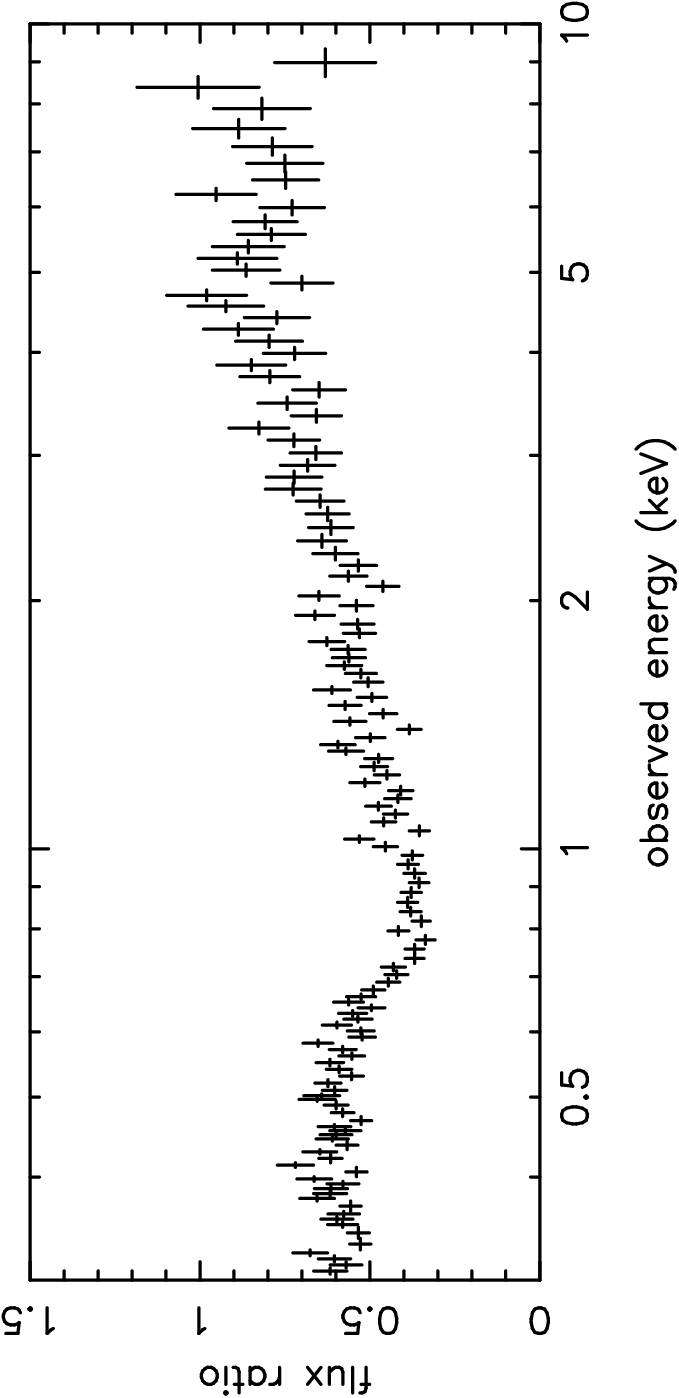}}
\rotatebox{-90}{
\epsscale{0.37}
\plotone{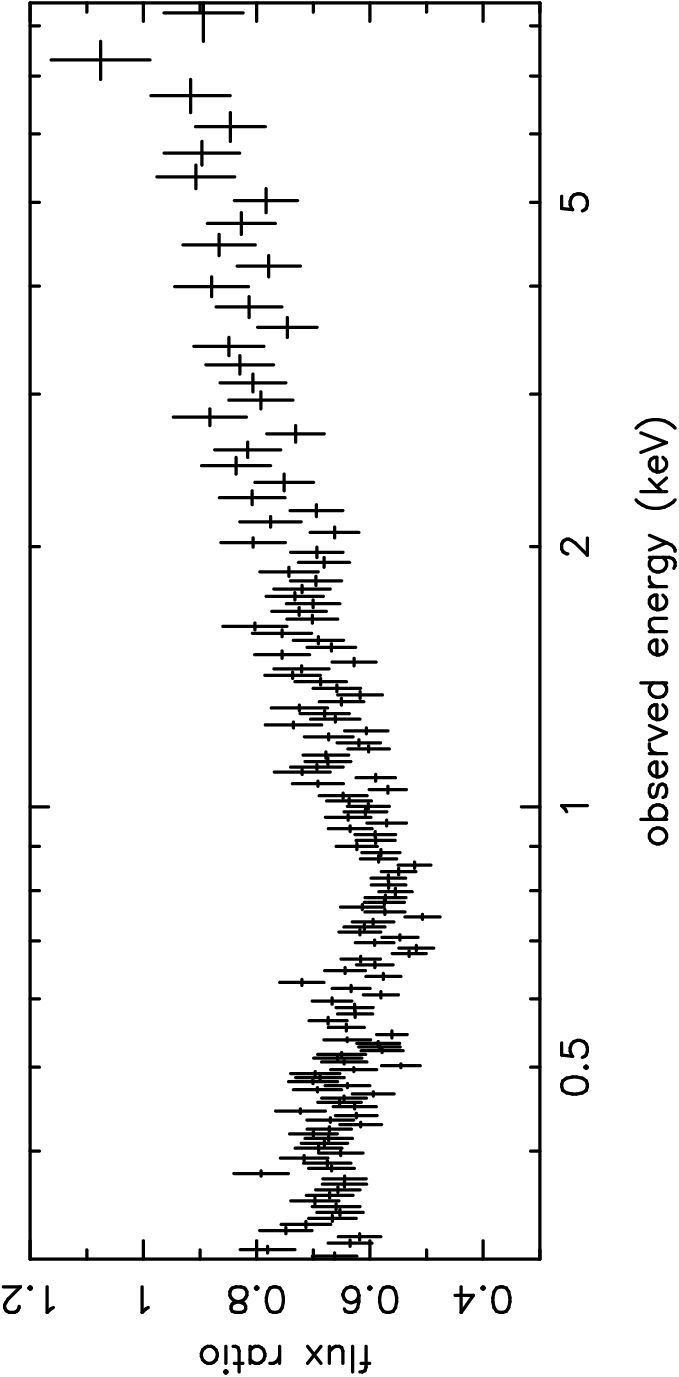}}
\caption{a.(top) pn camera data from the mid-low flux state observation divided by the data from the intermediate 
flux state
observation. b.(lower) pn camera data from the mid-high flux state observation divided by the data from the high 
flux state
observation.\label{fig9}}
\end{figure}

\clearpage

\begin{figure}
\rotatebox{-90}{
\epsscale{0.39}
\plotone{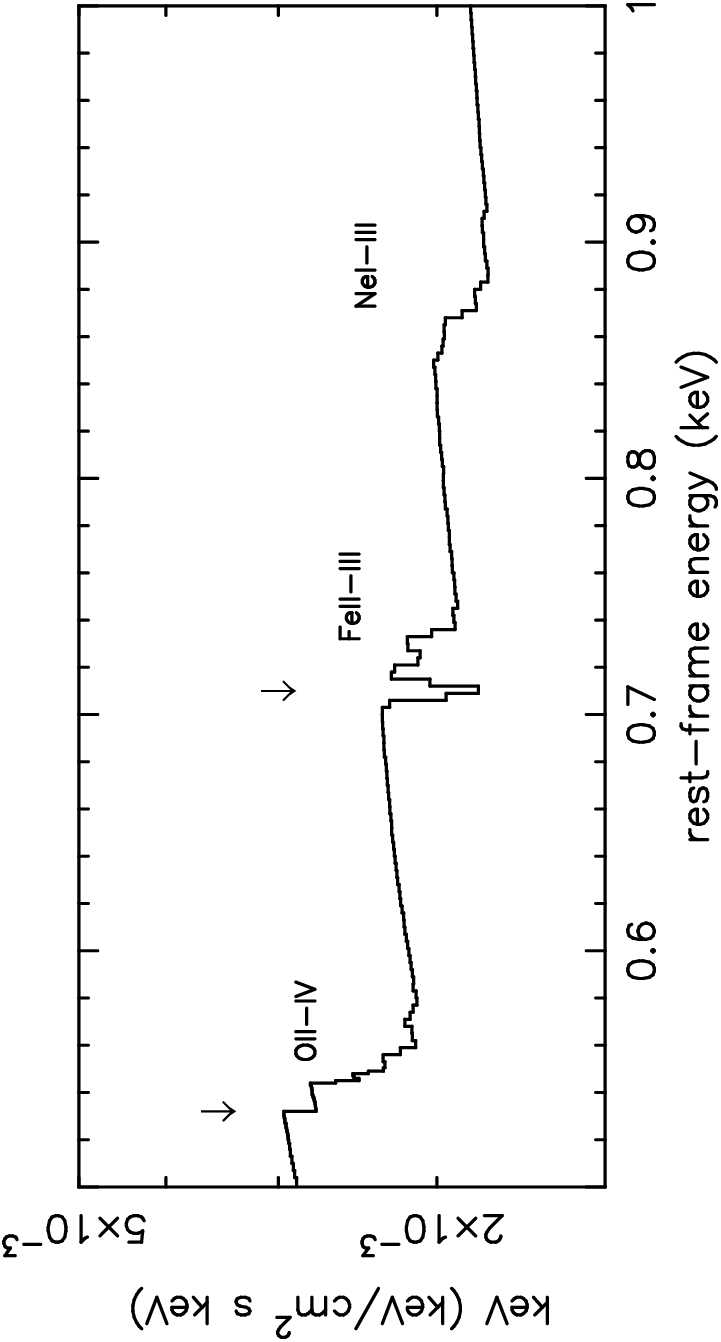}}
\rotatebox{-90}{
\epsscale{0.35}
\plotone{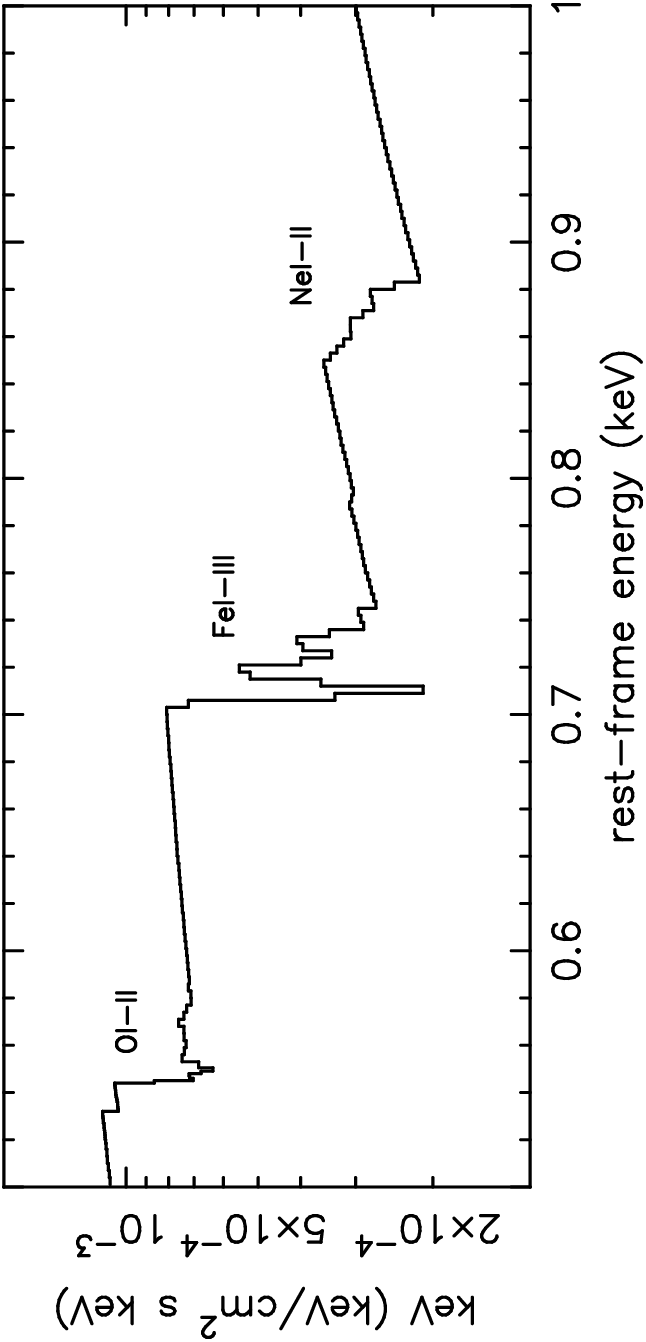}}
\caption{a.(top) Section of the compTT and XSTAR model fitted to the intermediate state difference spectrum 
showing the absorption 
edge structure in O, Fe, and Ne. The arrows note the OI edge due to the interstellar column in line-of-sight to 
\1h\ and
the Fe 2-3 UTA.  b.(lower) Same plot for the
corresponding fit to the mid-low state difference spectrum. Both plots are adjusted to the rest-frame of \1h.
\label{fig10}}
\end{figure}

\clearpage

\begin{figure}
\rotatebox{-90}{
\epsscale{0.7}
\plotone{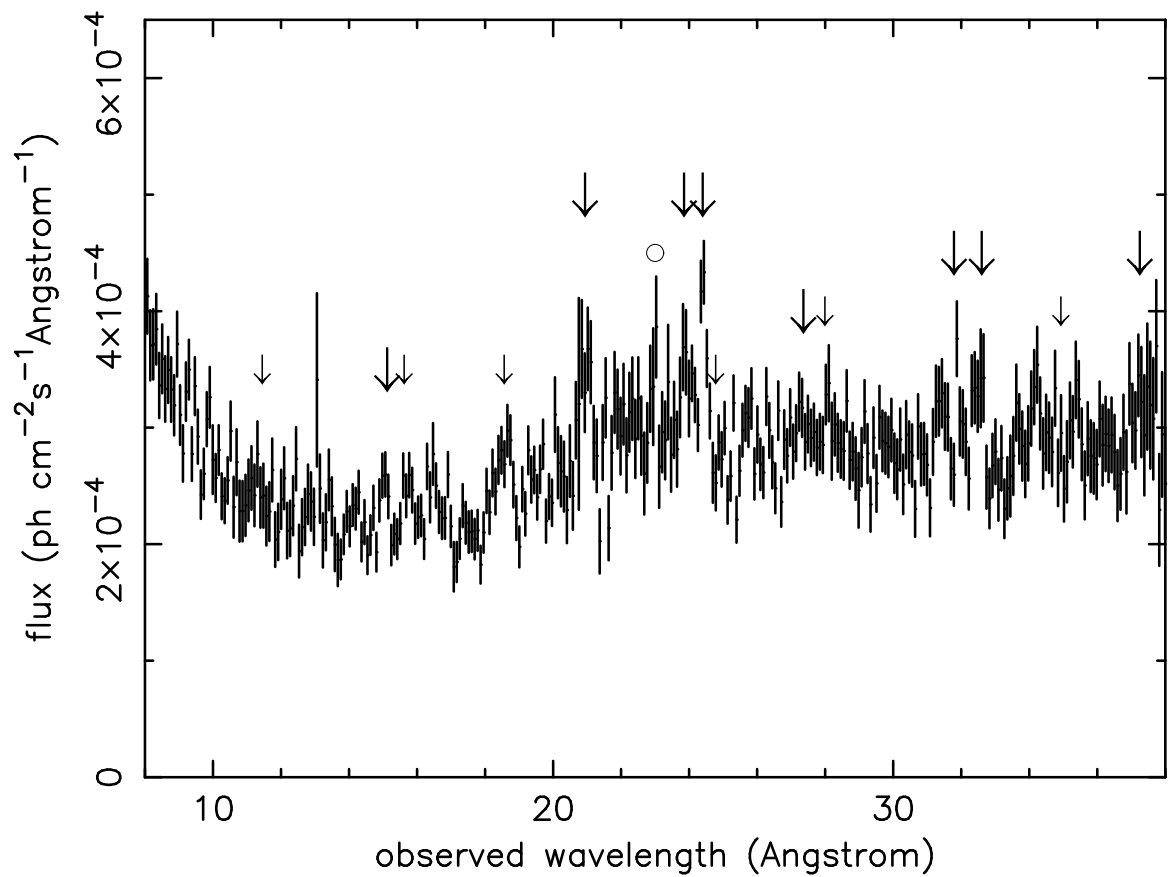}}
\caption{RGS spectrum of \1h\ summed over all 5 new \xmm\ observations. The fluxed spectrum is binned at 85 m\AA\ 
resolution. 
Reading 
from left to right the larger arrows indicate the 
wavelengths of the principal candidate emission lines of NeIX (f), OVIII Ly$\alpha$, OVII (r,f), NVII Ly$\alpha$, 
NVI (r,f) and CVI Ly$\alpha$, with 
the smaller arrows indicating
the threshold wavelengths of the RRC of NeIX, OVIII, OVII, NVI, CVI and CV. The open circle at $\sim$23 \AA\ 
notes a calibration defect.
\label{fig11}}
\end{figure}

\clearpage

\begin{figure}
\rotatebox{-90}{
\epsscale{0.36}
\plotone{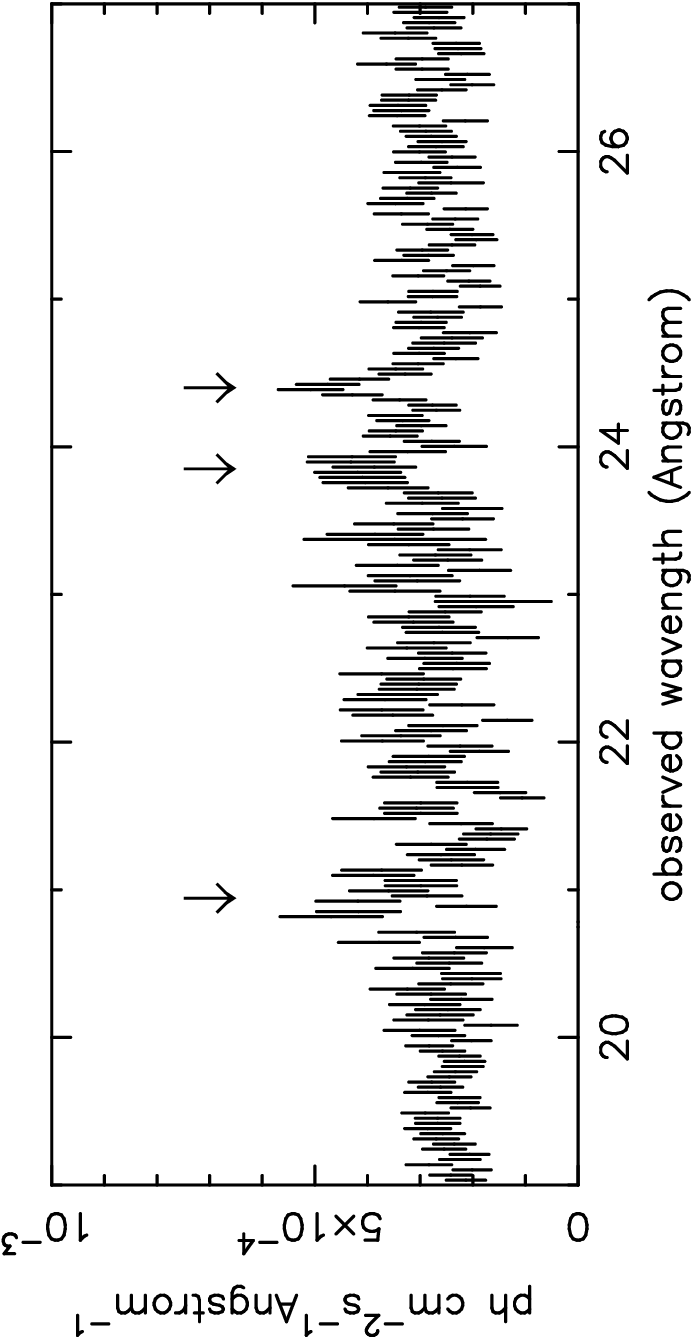}}
\rotatebox{-90}{
\epsscale{0.35}
\plotone{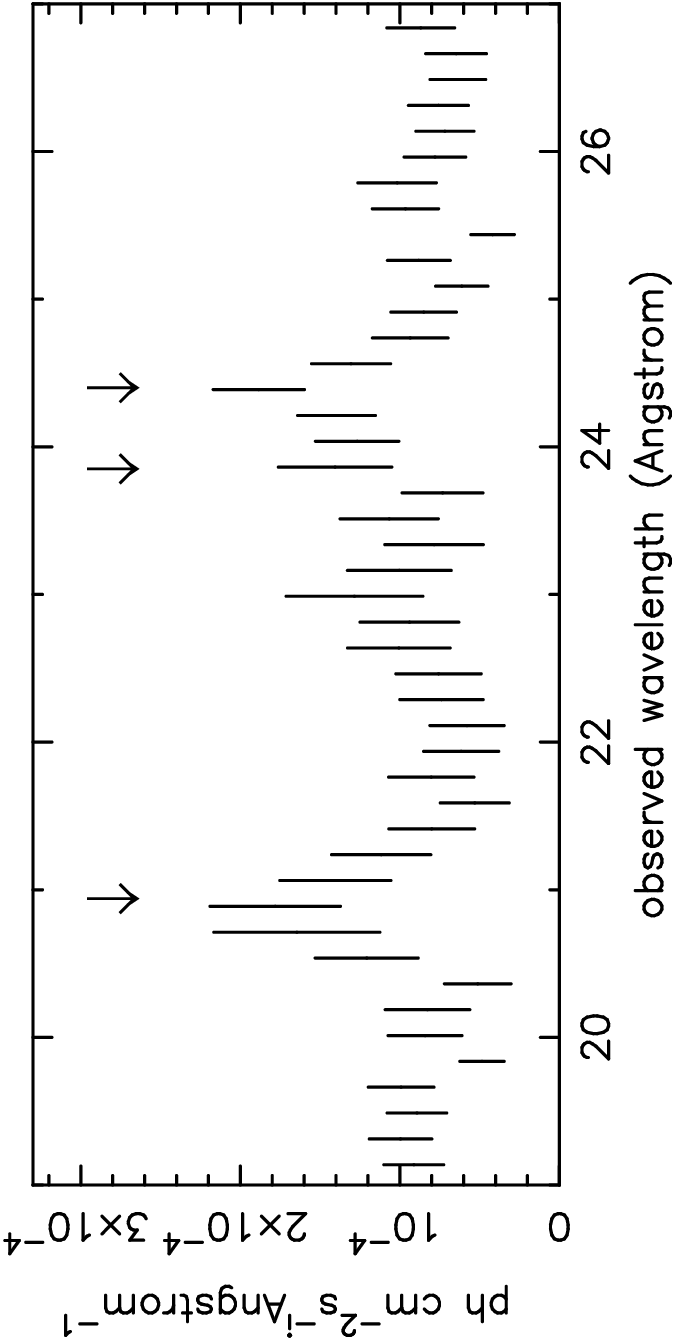}}
\caption{a.(top) Fluxed intermediate state RGS spectrum of \1h\ covering the waveband of OVIII Ly$\alpha$ and the 
OVII 1s-2p triplet binned at 
35 m\AA\ resolution. b.(lower) The same spectral band from the low
flux state observation of \1h\ binned at 170 m\AA\ resolution. The peak near 23 \AA\ is due to imperfect 
modelling of the 
O-K absorption edge in the detector response function. \label{fig12}}
\end{figure}

\clearpage

\begin{figure}
\rotatebox{-90}{
\epsscale{0.7}
\plotone{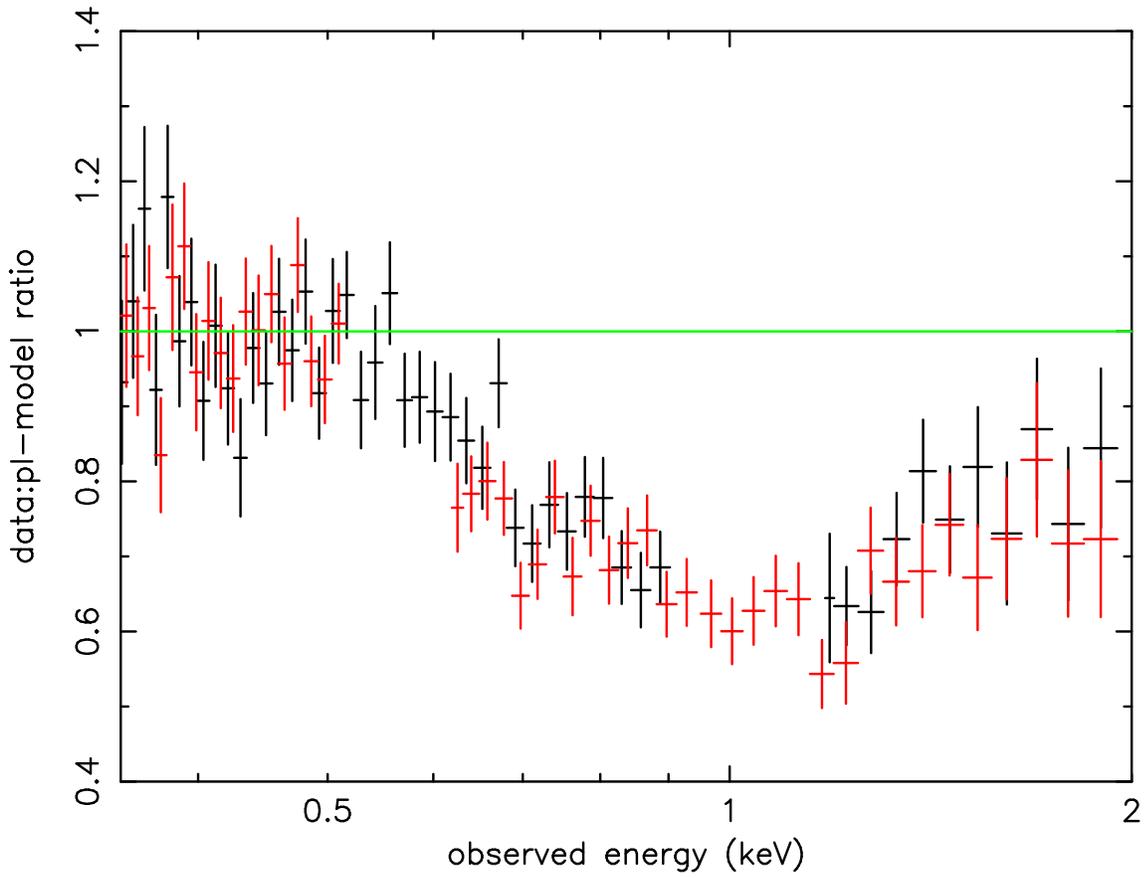}}
\caption{Coarsely binned RGS intermediate state difference spectrum of \1h\ plotted against a power law to 
illustrate absorption
edge structures consistent with those seen in the simultaneous EPIC spectrum \label{fig11}}
\end{figure}

\clearpage

\begin{figure}
\rotatebox{-90}{
\epsscale{0.24}
\plotone{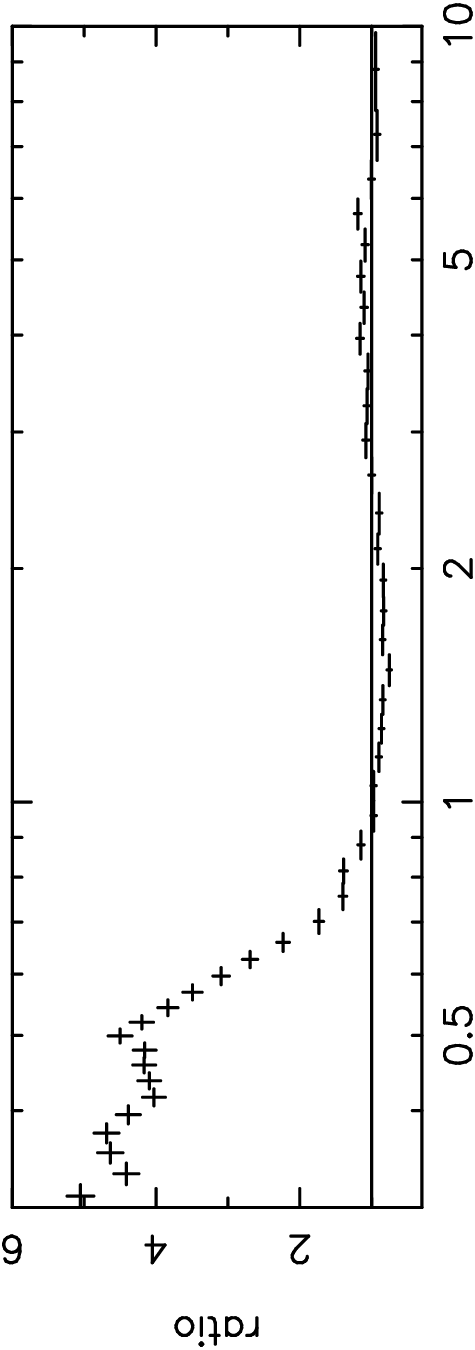}}
\rotatebox{-90}{
\epsscale{0.24}
\plotone{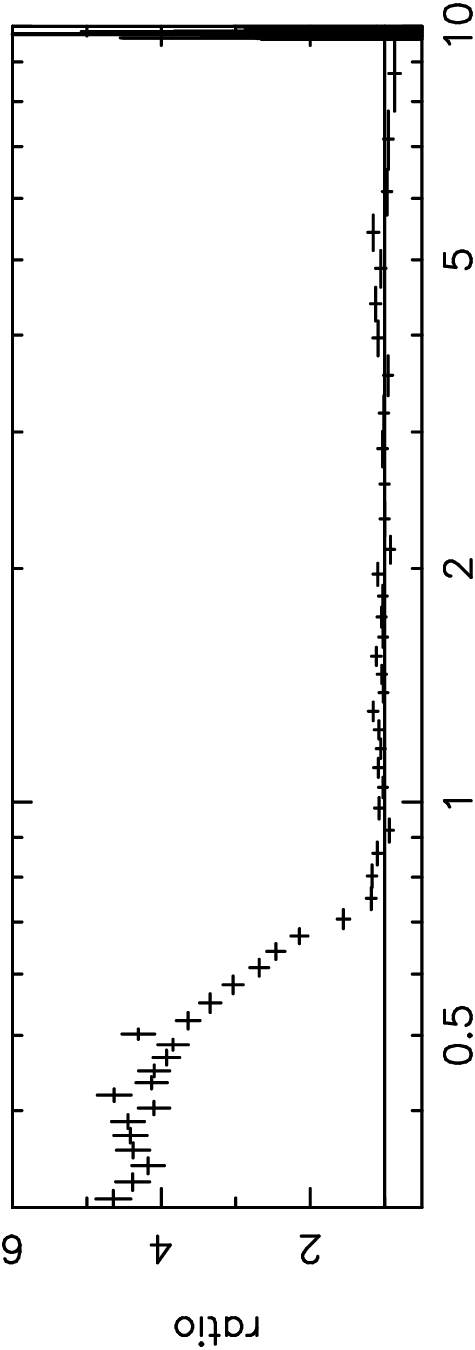}}
\rotatebox{-90}{
\epsscale{0.236}
\plotone{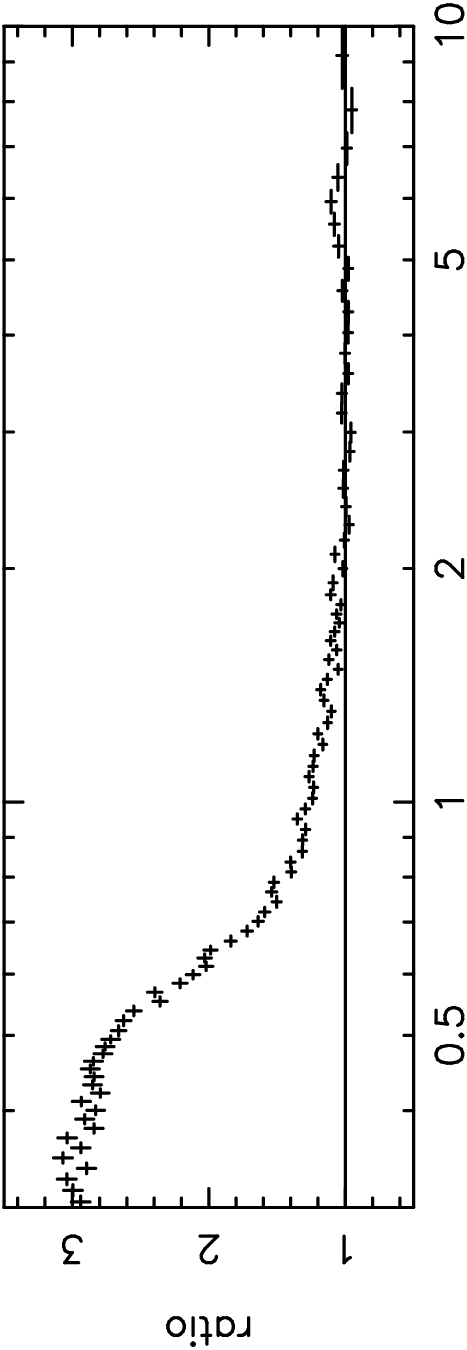}}
\rotatebox{-90}{
\epsscale{0.236}
\plotone{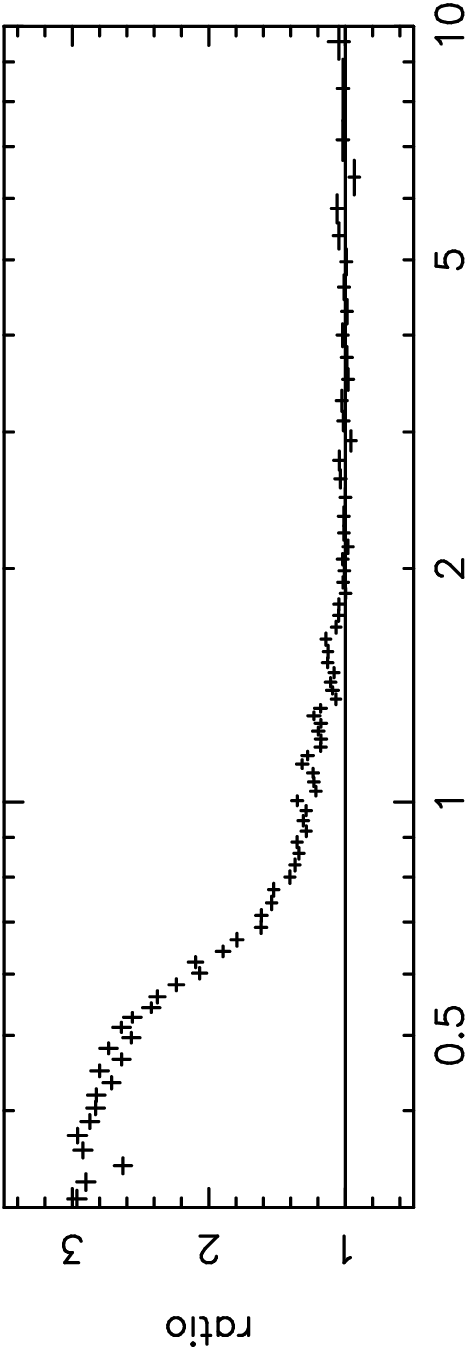}}
\rotatebox{-90}{
\epsscale{0.27}
\plotone{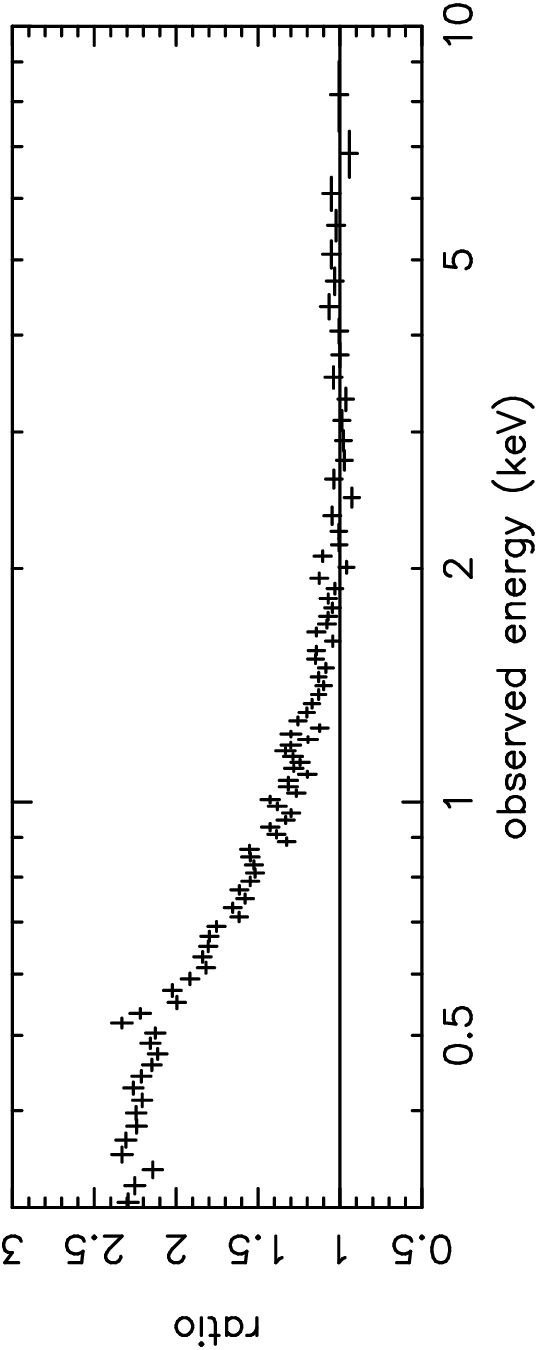}}
\caption{Reading from the top, the soft excess above a 2--10 keV power law fit for the low, mid-low, 
intermediate, mid-high and high
flux states of \1h. The respective 2-10 keV photon indices are: 1.07, 1.32, 1.61, 1.69 and 1.86
\label{fig14}}
\end{figure}

\clearpage

\begin{figure}
\rotatebox{-90}{
\epsscale{0.35}
\plotone{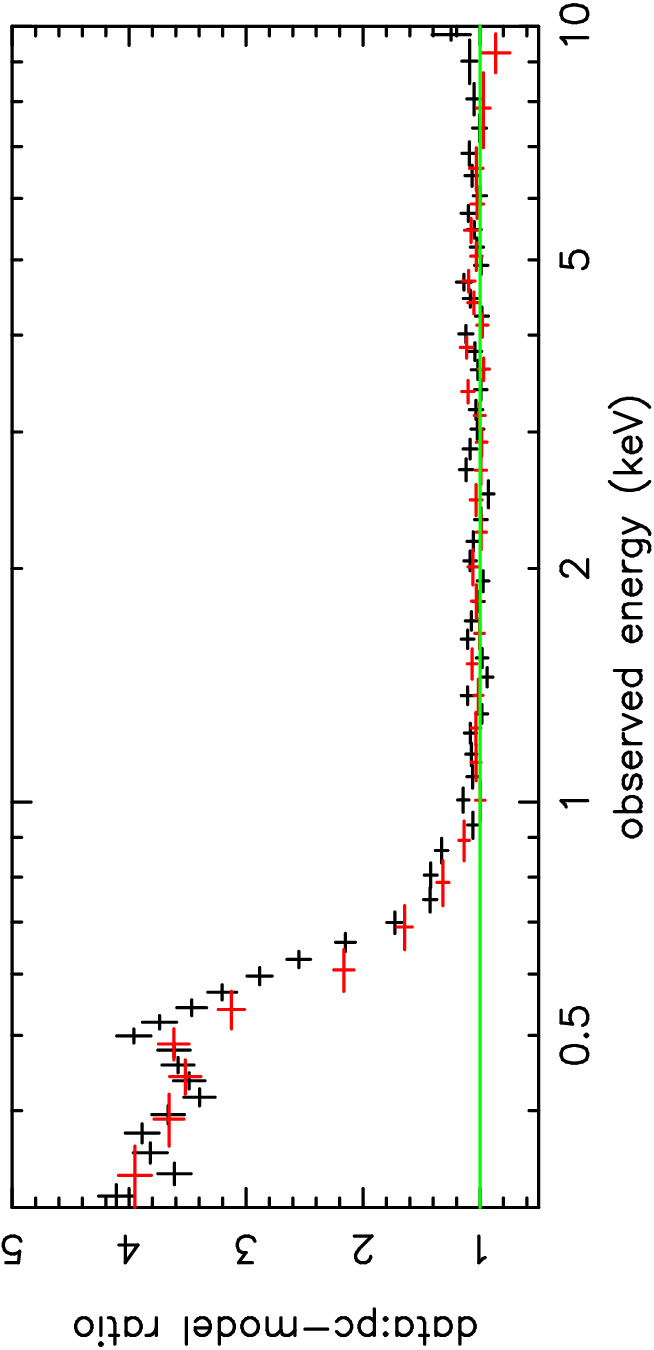}}
\rotatebox{-90}{
\epsscale{0.345}
\plotone{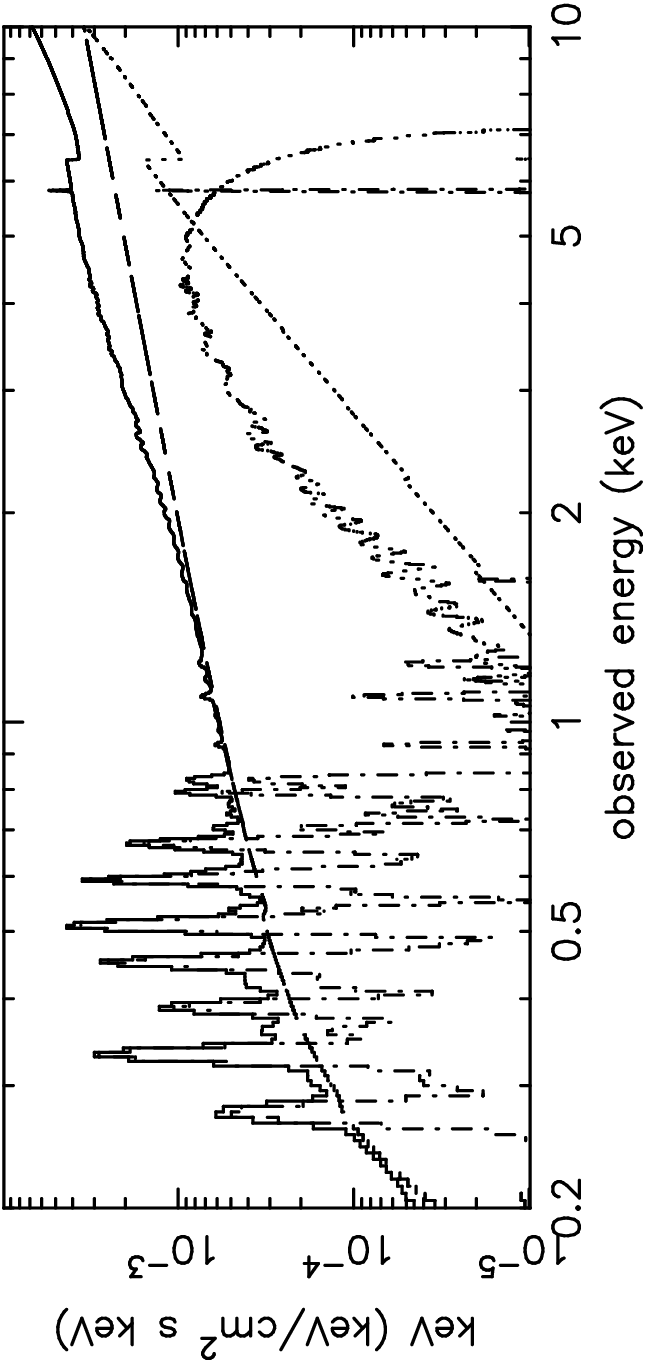}}
\caption{a.(top) Core soft excess obtained by subtracting the blackbody and 0.6 keV gaussian line components 
from the 0.3-10 keV
power law plus laor line fit to the low state EPIC data. b.(lower) XSTAR emission spectrum replacing the 
blackbody and gaussian
line components in an alternative fit to the core soft excess. Details 
are given in 
section 9. \label{fig15}}
\end{figure}



\clearpage

\begin{table}
\begin{center}
\caption{Identified emission features in the RGS spectrum}
\begin{tabular}{lccccc}
\tableline
\tableline
&\\
Feature & $\lambda$ (\AA) & $\sigma$/kT (eV) & Flux (10$^{-5}$ ph cm$^{-2}$ s$^{-1}$) & EW (eV) & $\Delta\chi^{2}$
\\ 
\tableline
O{\sc viii} Ly$\alpha$ & 19.0 & 5~$\pm$~2 & 6~$\pm$~2 & 9~$\pm$~3 & 22 \\
O{\sc vii} 1s-2p (r) & 21.6 & 5~$\pm$~2 & 6~$\pm$~2 & 7~$\pm$~2.5 & 22 \\
O{\sc vii} 1s-2p (f) & 22.1 & 1~$\pm$~1 & 4~$\pm$~1.5 & 3.5~$\pm$~1.2 & 32 \\
N{\sc vi} 1s-2p (f) & 29.5 & 1~$\pm$~1 & 5~$\pm$~3 & 3~$\pm$~1.5 & 14 \\
O{\sc vii} RRC & 16.8 & 3.3~$\pm$~1.8 & 1.9~$\pm$~1.1 & 4~$\pm$~2 & 18 \\

\tableline
\end{tabular}

\end{center}
\end{table}

\end{document}